\documentclass[aps,pre,twocolumn,longbibliography]{revtex4-1}

\usepackage{amssymb,amsmath}

\usepackage{graphicx}

\usepackage{color}

\newcommand\beq{\begin{equation}}
\newcommand\eeq{\end{equation}}
\newcommand\beqa{\begin{eqnarray}}
\newcommand\eeqa{\end{eqnarray}}
\newcommand{\dd}{\text{d}}

\newcommand{\al}{\alpha}

\begin{document}
\title{Enskog kinetic theory for a model of a confined quasi-two-dimensional granular fluid}
\author{Vicente Garz\'o}
\email{vicenteg@unex.es} \homepage{http://www.unex.es/eweb/fisteor/vicente/}
\affiliation{Departamento de
F\'{\i}sica and Instituto de Computaci\'on Cient\'{\i}fica Avanzada (ICCAEx), Universidad de Extremadura, E-06071 Badajoz, Spain}
\author{Ricardo Brito}
\affiliation{Departamento de Estructura de la Materia, F\'{\i}sica T\'ermica y Electr\'onica and GISC, Universidad Complutense de Madrid, Spain}
\author{Rodrigo Soto}
\affiliation{Departamento de
F\'{\i}sica, Facultad de Ciencias F\'{\i}sicas y Matem\'aticas, Universidad de Chile, Santiago, Chile}

\begin{abstract}
The Navier--Stokes transport coefficients for a model of a confined quasi-two-dimensional granular gas of smooth inelastic hard spheres are derived from the Enskog kinetic equation. A normal solution to this kinetic equation is obtained via the Chapman--Enskog method for states close to the {local}  homogeneous state. The analysis is performed to first order in spatial gradients, allowing the identification of the Navier--Stokes transport coefficients associated with the heat and momentum fluxes. The transport coefficients are determined from the solution to a set of coupled linear integral equations analogous to those for elastic collisions. These integral equations are solved by using the leading terms in a Sonine polynomial expansion. The results are particularized to the relevant state with stationary temperature, where explicit expressions for the Navier--Stokes transport coefficients are given in terms of the coefficient of restitution and the solid volume fraction. The present work extends to moderate densities previous results [Brey \emph{et al.} Phys. Rev. E \textbf{91}, 052201 (2015)] derived for low-density granular gases.
\end{abstract}

\date{\today}
\maketitle

\section{Introduction}
\label{sec0}

When granular media is externally excited (rapid flow conditions), the motion of grains is quite similar to the random motion of atoms or molecules of an ordinary fluid. Under these conditions, granular media flow like a fluid and kinetic theory can be employed to describe their dynamics. In contrast to ordinary fluids, the collisions between grains are inelastic and hence one has to inject energy into the system to sustain it under rapid flow conditions. In this context, granular fluids can be considered as complex systems that inherently are out of equilibrium \cite{JNB96,G03}. Their properties depend both on the dissipative collisions and also on the external energy mechanism used to drive the system.

The energy injection can be done either by driving through the boundaries, for example, shearing the system or vibrating its walls \cite{YHCMW02}, or alternatively by bulk driving, as in air-fluidized beds \cite{AD06,SGS05}. Nevertheless, one of the main problems in this sort of injection is that in most of the cases strong spatial gradients are generated in the system. This means that the theoretical description of these situations lies beyond the conventional Navier--Stokes domain (namely, when fluxes are linear functions of the spatial gradients). Thus, to avoid these difficulties, it is quite usual in theoretical and computational studies to heat the system by means of external driving forces \cite{thermostat}. In this driven case, the energy injected by the external force compensates on average the energy lost by collisions so that the system achieves a stationary nonequilibrium state. However, as expected, these forces do not play a neutral role in transport and hence, the Navier--Stokes transport coefficients obtained in the absence of them \cite{BDKS98,GD99,L05,GD02,GDH07} are different from those derived in the presence of the external forces \cite{GM02,GMT13,GChV13,KG15}.

However, an alternative to the use of external forces has gained interest in the past few years \cite{OU98,PMEU04,M05,C08,PCR09,RPGRSCM11,GSVP11,CMS12,PGGSV12}. The idea is to use a particular geometry (quasi-two-dimensional geometry) where the energy is placed into the bulk region and generates homogeneous states. In this geometry the granular gas is confined in a box that is very large in the horizontal directions but its vertical length is smaller than two particle diameters. Under these conditions, since grains cannot be on top one each other, the system can be seen as a monolayer. When the box is vertically vibrated, energy is injected into the vertical degrees of freedom of particles via the collisions of grains with the top and bottom plates. This energy gained by collisions with the walls is then transferred to the horizontal degrees of freedom by collisions between grains. When the system is observed from above, it is fluidized and can remain homogeneous in a wide range of parameters \cite{PMEU04,C08}.

A model accounting for the transfer of energy from the vertical to horizontal degrees of freedom in the quasi-two-dimensional geometry has been recently proposed by Brito \emph{et al.} \cite{BRS13}. As in the case of the conventional smooth inelastic hard sphere (IHS) model, the inelasticity of collisions is characterized by a constant (positive) coefficient of normal restitution $\al\leq 1$. In addition, an extra velocity $\Delta$ is added to the relative motion of colliding particles so that the magnitude of the normal component of the relative velocity is increased by a factor $2\Delta$ in the collision. This term models the energy transfer from the vertical degrees of freedom to the horizontal degrees of freedom.

The above simple extension of the  IHS model presents several properties that makes it relevant to describe the experiments and useful as a theoretical model. The most important is that, as it is the case for experiments, the energy transfer occurs only at collisions, conserving horizontal momentum. Hence, at the macroscopic level, a hydrodynamic description is expected, where only the energy equation is modified by the
addition of a sink/source term \cite{BRS13}. The system can be placed arbitrarily close to equilibrium by taking simultaneously the
limits $\Delta\to0$ and $\al\to1$, allowing to make perturbation analysis and, also, far from equilibrium conditions can be studied
when small values of $\al$ are used. Quite generally, homogeneous stable states are possible to generate for this collisional model,
allowing to test the predictions of kinetic theory  and to compare with experiments for homogeneous systems in all ranges of $\al$.
The magnitude of $\Delta$ is related to the intensity of the vertical vibrations. Although the energy transfer depends on several factors,
like the impact parameter in the vertical collisions, or the total energy stored in the $z$-direction, we assume, for simplicity, a constant value of $\Delta$, which scales with the velocity of the vibrating walls. This election has the advantage of introducing only a single additional parameter to the IHS model and additionally, the system remains homogeneous for all densities \cite{BRS13}.

There are other collision rules that model the vertical-to-horizontal energy transfer. For instance, Ref.\ \cite{GS} considers an stochastic coefficient of restitution, that can take values larger or smaller than 1. However, it lacks an energy scale, and therefore it is not a suitable model for a vertically vibrated case. Recently, it has been proposed a more realistic model, where the $\Delta$ parameter depends on the local density \cite{RissoSotoGuzman}. Such model gives rise to a van der Waals loop and a phase separation, in agreement with experiments \cite{C08}. However,  this collisional model is much more involved to derive the hydrodynamic equations as an additional field is needed. It is also possible to envisage other extensions, for example, for frictional grains with rotational degrees of freedom, the tangential relative velocity can also be increased at collisions.

The collisional model with constant $\Delta$ (referred to as the Delta-model) has been widely studied by Brey and co-workers in several papers: the study includes the homogeneous state \cite{BGMB13,BMGB14,BGMB14} as well as the determination of the Navier--Stokes transport coefficients for a low-density granular gas \cite{BBMG15}. An independent calculation for the shear viscosity of a dilute granular gas was carried out by Soto \emph{et al.} \cite{SRB14} together with computer simulations. Comparison between them  shows good quantitative agreement for strong values of inelasticity (say for instance, $\al \lesssim 0.8$).

The objective of this paper is to extend the previous efforts made for the Delta-model \cite{BBMG15,SRB14} to the revised Enskog theory (RET) for a description of hydrodynamics and transport for higher densities. It is known that the RET for elastic collisions \cite{BE73} gives an accurate kinetic theory over the entire fluid domain. In the case of inelastic collisions, theoretical results \cite{GD99,G05} derived from the RET have shown to be in good agreement with molecular dynamics (MD) simulations \cite{MDCPH11,MGHEH12,MGH14} and even with real experiments \cite{YHCMW02}. This confirms the reliability of the RET for describing granular flows in conditions of practical interest.

As for the IHS model, the Chapman--Enskog method \cite{CC70} conveniently adapted for dissipative dynamics will be employed to solve the Enskog kinetic equation and obtain the Navier--Stokes hydrodynamics and the associated transport coefficients. As expected, the analysis here provides formally exact results for the Navier--Stokes transport coefficients in terms of the solutions to linear integral equations. These equations are approximately solved by considering the leading terms in a Sonine polynomial expansion. However, given the technical difficulties embodied in the calculation of the transport coefficients in the time-dependent problem,  here the relevant state of a confined granular fluid at the state with stationary temperature is mainly considered. This allows us to express the (scaled) transport coefficients in terms of the coefficient of restitution and the solid volume fraction.

There are several motivations to extend previous works \cite{BGMB14,SRB14} to moderately dense gases. First, by extending the Boltzmann analysis to high densities comparison with MD simulations becomes practical. For instance, a comparison of the dependence of both density and coefficient of restitution on the theoretical shear viscosity with that from MD could determine the validity of the kinetic theory of the Delta-model. Second, accurate predictions from the RET could be also compared against experimental data performed for the quasi-two-dimensional geometry mentions before. Therefore, the results reported in this paper provide the basis for practical quantitative applications of the Delta-model.
Finally, the Delta-model can be considered also as a thermostated granular model that can be easily implemented, for which it is relevant to understand its properties in the full range of densities.

The plan of the paper is as follows. The Delta-model is introduced first in Sec.\ \ref{sec1}, summarizing its main properties. Then, the Enskog kinetic equation for the model is displayed, and the exact balance equations for the densities of mass, momentum, and energy are derived from it. This allows us to express the cooling rate and the kinetic and collisional transfer contributions to the fluxes in terms of the velocity distribution function. Section \ref{sec2} deals with the application of the Chapman--Enskog method to the Enskog equation. The results for the momentum and heat fluxes to first order in the spatial gradients are provided, with some details of the calculation appearing in two Appendices. The Navier--Stokes transport coefficients are formally obtained in Sec.\ \ref{sec3} as the solutions of a set of coupled linear inhomogeneous integral equations. These integral equations are explicitly solved at the stationary temperature state in Sec.\ \ref{sec4} by assuming for simplicity a Gaussian distribution for the zeroth-order approximation. The explicit expressions of the relevant (scaled) transport coefficients are displayed in Table \ref{table1} for a two-dimensional system as functions of the density and the coefficient of restitution. The paper is closed in Sec.\ \ref{sec5} with a brief discussion of the results.


\begin{table*}[tbp]
\caption{Summary of the main results of the paper at the stationary temperature. Two-dimensional granular gas ($d=2$).}
\label{table1}
\begin{ruledtabular}
\begin{tabular}{c}
\vspace{0.2cm}
$\eta^*=\left[1+\frac{1}{2}\phi \chi \left(1+\alpha+\sqrt{\frac{2}{\pi}}\Delta_\text{s}^*\right)\right]\eta_\text{k}^*
+\frac{1}{2}\gamma^*$,\\
\vspace{0.2cm}
$\eta_\text{k}^*=\nu_\eta^{^*-1}\left\{1-\frac{1}{4}\phi \chi\left[(1+\al)
(1-3 \alpha) -4\sqrt{\frac{2}{\pi}}(1+2\al)\Delta_\text{s}^*-4\Delta_\text{s}^{*2}\right]\right\}$,\\
\vspace{0.2cm}
$\gamma^*=\frac{8}{\pi}\phi^2 \chi \left(1+\alpha+\sqrt{\frac{\pi}{2}}\Delta_\text{s}^*\right)$,\\
\vspace{0.2cm}
$\kappa^*=\left[1+\frac{3}{4}\phi \chi \left( 1+\alpha+\sqrt{\frac{2}{\pi}}
\Delta_\text{s}^*\right)\right]\kappa_\text{k}^*
+\frac{2}{\pi}\phi^2 \chi \left(1+\alpha+\sqrt{\frac{\pi}{2}}
\Delta_\text{s}^*\right)$,\\
\vspace{0.2cm}
$\kappa_\text{k}^*=\frac{1}{2\nu_\kappa^*+\Delta_s^*\Big(\frac{\partial \zeta_0^*}{\partial \Delta^*}\Big)_s}
\left\{1+\frac{3}{8}\phi \chi(1+\alpha)^2(2\alpha-1)-\frac{\Delta_\text{s}^*}{\sqrt{2\pi}}\phi \chi \left[\frac{3}{4}+3(1+\al)\left(1-\sqrt{\frac{\pi}{2}}\Delta_{\text{s}}^*\right)
-\frac{9}{2}(1+\al)^2-\Delta_{\text{s}}^{*2}\right]\right\}$,\\
\vspace{0.2cm}
$\mu^*=\left[1+\frac{3}{4}\phi \chi \left( 1+\alpha+\sqrt{\frac{2}{\pi}}
\Delta_\text{s}^*\right)\right]\mu_\text{k}^*$,\\
\vspace{0.2cm}
$\mu_\text{k}^*=-\frac{1}{\nu_\kappa^*}\phi \chi \left(1+\frac{1}{2}\phi\partial_\phi\ln \chi\right)\left\{\frac{3}{8}
\al (1-\al^2)-\frac{\Delta_\text{s}^*}{\sqrt{2\pi}}\phi \chi \left[2\Delta_\text{s}^{*2}-3\left(\frac{1}{2}-\al^2\right)
+\frac{3}{2}\sqrt{2\pi}\al\Delta_\text{s}^*\right]\right\}$,\\
\vspace{0.2cm}
$\nu_\eta^*=\frac{3}{8}\chi \left[\left(\frac{7}{3}-\alpha\right)(1+\alpha)
+\frac{2\sqrt{2\pi}}{3}(1-\al)\Delta_\text{s}^*-\frac{2}{3}\Delta_\text{s}^{*2}\right]$,\\
\vspace{0.2cm}
$\nu_\kappa^*=\nu_\mu^*=\frac{1+\alpha}{2}\chi\left[
\frac{1}{2}+\frac{15}{8}(1-\alpha)\right]-\frac{\Delta_\text{s}^*}{16}\chi
\left[\sqrt{2\pi}(5\al-1)+10\Delta_\text{s}^{*}\right]$,\\
\vspace{0.2cm}
$\Delta_\text{s}^*(\al)=\frac{1}{2}\sqrt{\frac{\pi}{2}}\al\left[\sqrt{1+
\frac{4(1-\al^2)}{\pi \al^2}}-1\right]$,\\
\vspace{0.2cm}
$p^*=1+\phi \chi (1+\al)+2\sqrt{\frac{2}{\pi}}\phi \chi\Delta_\text{s}^*$,\\
\vspace{0.2cm}
$\chi=\frac{1-\frac{7}{16}\phi}{(1-\phi)^2}$.
\end{tabular}
\end{ruledtabular}
\end{table*}

\section{Enskog kinetic equation. The collisional model}
\label{sec1}

\subsection{Collisional model}

We consider a granular fluid modeled as a gas of smooth inelastic hard spheres of mass $m$ and diameter $\sigma$. Let $(\mathbf{v}_1, \mathbf{v}_2)$ denote the pre-collisional velocities of two spherical particles. The collision rule for the post-collisional velocities $(\mathbf{v}_1',\mathbf{v}_2')$ reads \cite{BRS13}
\begin{align}
\mathbf{v}_1'&=\mathbf{v}_1-\frac{1}{2}\left(1+\alpha\right)(\widehat{{\boldsymbol {\sigma }}}\cdot \mathbf{g})\widehat{{\boldsymbol {\sigma }}}-\Delta \widehat{{\boldsymbol {\sigma }}},\label{1.1}\\
{\bf v}_{2}'&=\mathbf{v}_{2}+\frac{1}{2}\left(1+\alpha\right)
(\widehat{{\boldsymbol {\sigma}}}\cdot \mathbf{g})
\widehat{\boldsymbol {\sigma}}+\Delta \widehat{{\boldsymbol {\sigma }}},\label{1.2}
\end{align}
where $\mathbf{g}=\mathbf{v}_1-\mathbf{v}_2$ is the relative velocity, $\widehat{{\boldsymbol {\sigma}}}$ is the unit collision vector joining
the centers of the two colliding spheres and pointing from particle 1 to particle 2, and particles are approaching if $\widehat{{\boldsymbol {\sigma}}}\cdot \mathbf{g}>0$. In Eqs.\ \eqref{1.1} and \eqref{1.2}, $\al$ is the (constant) coefficient of normal restitution defined in the interval $0<\alpha\leq 1$, and $\Delta$ is an extra velocity added to the relative motion. This extra velocity points outward in the normal direction $\widehat{\boldsymbol {\sigma}}$, as required by the conservation of angular momentum \cite{L04}. The relative velocity after collision is
\beq
\label{1.3}
\mathbf{g}'=\mathbf{v}_1'-\mathbf{v}_2'=\mathbf{g}-(1+\al)(\widehat{{\boldsymbol {\sigma}}}\cdot \mathbf{g})
\widehat{\boldsymbol {\sigma}}-2\Delta \widehat{{\boldsymbol {\sigma }}},
\eeq
so that
\beq
\label{1.4}
(\widehat{{\boldsymbol {\sigma}}}\cdot \mathbf{g}')=-\al (\widehat{{\boldsymbol {\sigma}}}\cdot \mathbf{g})-2\Delta.
\eeq

With the set of collision rules \eqref{1.1} and \eqref{1.2}, momentum is conserved but energy is not. The change in kinetic energy upon collision is
\begin{align}
\label{1.5}
\Delta E&\equiv \frac{m}{2}\left(v_1^{'2}+v_2^{'2}-v_1^2-v_2^2\right)\nonumber\\
&=m\left[\Delta^2+\al \Delta (\widehat{{\boldsymbol {\sigma}}}\cdot \mathbf{g})-\frac{1-\al^2}{4}(\widehat{{\boldsymbol {\sigma}}}\cdot \mathbf{g})^2\right].
\end{align}
The right-hand side of Eq.\ \eqref{1.5} vanishes for elastic collisions ($\al=1$) and $\Delta=0$. Moreover, it appears that energy can be gained or lost in a collision depending on whether $\widehat{{\boldsymbol {\sigma}}}\cdot \mathbf{g}$ is smaller than or larger than $2\Delta /(1-\al)$.

For practical purposes, it is also convenient to consider the restituting collision $\left(\mathbf{v}_1'',\mathbf{v}_2''\right)\to \left(\mathbf{v}_1,\mathbf{v}_2\right)$ with the same collision vector $\widehat{{\boldsymbol {\sigma }}}$:
\begin{align}
\mathbf{v}_1''=\mathbf{v}_1-\frac{1}{2}\left(1+\alpha^{-1}\right)(\widehat{{\boldsymbol {\sigma }}}\cdot \mathbf{g})\widehat{{\boldsymbol {\sigma }}}-\alpha^{-1}\Delta \widehat{{\boldsymbol {\sigma }}},\label{1.6}\\
\mathbf{v}_2''=\mathbf{v}_2+\frac{1}{2}\left(1+\alpha^{-1}\right)(\widehat{{\boldsymbol {\sigma }}}\cdot \mathbf{g})\widehat{{\boldsymbol {\sigma }}}+\alpha^{-1}\Delta \widehat{{\boldsymbol {\sigma }}}. \label{1.7}
\end{align}
The volume transformation in velocity space for a direct collision is $\dd \mathbf{v}_1'\dd \mathbf{v}_2'=\al \dd \mathbf{v}_1\dd \mathbf{v}_2$, and for the restituting collision is
$\dd \mathbf{v}_1''\dd \mathbf{v}_2''=\al^{-1} \dd \mathbf{v}_1\dd \mathbf{v}_2$.

\subsection{Enskog kinetic equation}

At a kinetic level, all the relevant information on the state of the system is provided by the one-particle velocity distribution function $f(\mathbf{r}, \mathbf{v}, t)$. For moderate densities, in the absence of external forces, the distribution $f(\mathbf{r}, \mathbf{v}, t)$ of the collisional model obeys the Enskog kinetic equation \cite{BGMB13}
\beq
\label{1.8}
\frac{\partial}{\partial t}f(\mathbf{r}, \mathbf{v},t)+\mathbf{v}\cdot \nabla f(\mathbf{r},\mathbf{v},t)=J_\text{E}[\mathbf{r},\mathbf{v}|f,f],
\eeq
where the Enskog collision operator $J_\text{E}$ of the model reads
\beqa
\label{1.9a}
& & J_\text{E}[\mathbf{r},\mathbf{v}_1|f,f]\equiv \sigma^{d-1}\int\dd{\bf v}_{2}\int \dd \widehat{\boldsymbol{\sigma}}
\Theta (-\widehat{{\boldsymbol {\sigma }}}\cdot {\bf g}-2\Delta)
\nonumber\\
& & \times(-\widehat{\boldsymbol {\sigma }}\cdot {\bf g}-2\Delta)
\al^{-2}\chi(\mathbf{r},\mathbf{r}+\boldsymbol{\sigma}) f(\mathbf{r},\mathbf{v}_1'',t)\nonumber\\
& & \times
f(\mathbf{r}+\boldsymbol{\sigma},\mathbf{v}_2'',t)
-\sigma^{d-1}\int\ \dd{\bf v}_{2}\int \dd\widehat{\boldsymbol{\sigma}}
\Theta (\widehat{{\boldsymbol {\sigma }}}\cdot {\bf g})
\nonumber\\
& & \times (\widehat{\boldsymbol {\sigma }}\cdot {\bf g})
\chi(\mathbf{r},\mathbf{r}+\boldsymbol{\sigma}) f(\mathbf{r},\mathbf{v}_1,t)
f(\mathbf{r}+\boldsymbol{\sigma},\mathbf{v}_2,t).
\eeqa
Like the Boltzmann equation, the Enskog equation neglects velocity correlations among particles that are about to collide, but it accounts for excluded volume effects and spatial correlations via the pair distribution function at contact $\chi(\mathbf{r},\mathbf{r}+\boldsymbol{\sigma})$. In Eq.\ \eqref{1.9a},  $\Theta(x)$ is the Heaviside step function and  $d$ is the dimensionality of the system. Note that although the system considered is two-dimensional, the calculations worked out here will be performed for an arbitrary number of dimensions $d$.

An important property of the integrals involving the Enskog collision operator is \cite{BGMB13,SRB14}
\begin{align}
\label{1.9}
I_\psi&\equiv \int\; \dd \mathbf{v}_1\; \psi(\mathbf{v}_1) J_\text{E}[\mathbf{r},\mathbf{v}_1|f,f]\nonumber\\
&=\sigma^{d-1}\int \dd \, \mathbf{v}_1\int\ \dd{\bf v}_{2}\int \dd\widehat{\boldsymbol{\sigma}}\,
\Theta (\widehat{{\boldsymbol {\sigma }}}\cdot {\bf g})(\widehat{\boldsymbol {\sigma }}\cdot {\bf g})
\nonumber\\
&\times
\chi(\mathbf{r},\mathbf{r}+\boldsymbol{\sigma}) f(\mathbf{r},\mathbf{v}_1,t)
f(\mathbf{r}+\boldsymbol{\sigma},\mathbf{v}_2,t)
\left[\psi(\mathbf{v}_1')-\psi(\mathbf{v}_1)\right],
\nonumber\\
\end{align}
where $\mathbf{v}_1'$ is defined by Eq.\ \eqref{1.1}. This is the same result as for the IHS model \cite{BP04}. A consequence of the relation \eqref{1.9} is that the balance equations of the densities of mass, momentum and energy can be derived by following similar mathematical steps as those made for the IHS model and adopt the standard form for rapid granular flows \cite{GD99,L05,GDH07}. They are given by
\begin{gather}
\label{1.10}
D_t n+n\nabla \cdot \mathbf{U}=0,\\
\label{1.11}
\rho D_t U_i+\partial_j P_{ij}=0,\\
\label{1.12}
D_t T+\frac{2}{dn}\left(\partial_i q_i+ P_{ij}\partial_j U_i \right)=-\zeta T,
\end{gather}
where $\rho=mn$ is the mass density, $\mathbf{U}$ is the velocity field, and $T$ is the granular temperature:
\begin{align}
n({\bf r},t) &= \int  \dd\mathbf{v} f({\bf r},{\bf v},t),\\
\mathbf{U}({\bf r},t) &= \frac{1}{n({\bf r},t) }\int  \dd\mathbf{v} {\bf v} f({\bf r},{\bf v},t),\\
\frac{d}{2}n({\bf r},t) T({\bf r},t) &= \int  \dd\mathbf{v} \frac{m}{2}{\bf V}^2 f({\bf r},{\bf v},t),
\end{align}
$\mathbf{V}=\mathbf{v}-\mathbf{U}$ being the peculiar velocity. In addition,  $D_t\equiv \partial_t+\mathbf{U}\cdot \nabla$ is the material derivative. The
cooling rate $\zeta$ is due to dissipative collisions, but contrary to the IHS model where it is always positive, it can take negative values for small temperatures [see Eq.~\eqref{1.17} below]. This property allows the system to reach stable steady states. The pressure tensor
$\mathsf{P}({\bf r},t)$ and
the heat flux $\mathbf{q}({\bf r},t)$ have both {\em kinetic} and {\em collisional transfer} contributions, i.e., ${\sf P}={\sf P}_\text{k}+{\sf P}_\text{c}$ and ${\bf q}={\bf q}_\text{k}+{\bf q}_\text{c}$. The kinetic contributions
are given as usual by
\begin{align}
\label{1.13}
{\sf P}_\text{k}({\bf r}, t)&=\int \; \dd{\bf v} m{\bf V}{\bf V}f({\bf r},{\bf v},t),\\
\label{1.13.1}
{\bf q}_\text{k}({\bf r}, t)&=\int \; \dd{\bf v} \frac{m}{2}V^2{\bf V}f({\bf r},{\bf v},t).
\end{align}
The collisional transfer contributions are (see the Appendix~\ref{appC} for details)
\beqa
\label{1.14}
\mathsf{P}_\text{c}&=&\frac{1+\alpha}{4}m \sigma^{d}
\int \dd\mathbf{v}_{1}\int \dd\mathbf{v}_{2}\int
\dd\widehat{\boldsymbol {\sigma }}\Theta (\widehat{\boldsymbol
{\sigma }}\cdot
\mathbf{g})(\widehat{\boldsymbol {\sigma}}\cdot \mathbf{g})
\nonumber\\
& \times &  \widehat{\boldsymbol {\sigma}}\widehat{\boldsymbol {\sigma }}\left[(\widehat{\boldsymbol {\sigma}}\cdot \mathbf{g})+\frac{2\Delta}{1+\al}\right]\nonumber\\
& \times &
\int_{0}^{1} \dd \lambda f_2\left[\mathbf{r}-\lambda{\boldsymbol{\sigma}},\mathbf{r}+(1-\lambda)
{\boldsymbol {\sigma}},\mathbf{v}_{1},\mathbf{v}_{2},t\right],
\nonumber\\
\eeqa
\beqa
\label{1.15}
{\bf q}_\text{c}&=&\frac{1+\alpha}{4}m \sigma^{d}
\int \dd\mathbf{v}_{1}\int \dd\mathbf{v}_{2}\int
\dd\widehat{\boldsymbol {\sigma}}\Theta (\widehat{\boldsymbol{\sigma}}\cdot
\mathbf{g})(\widehat{\boldsymbol {\sigma}}\cdot \mathbf{g})^{2}
\nonumber\\
&\times&
(\widehat{\boldsymbol {\sigma}}\cdot {\bf G})
\widehat{\boldsymbol {\sigma}}\int_{0}^{1}\dd \lambda f_2\left[\mathbf{r}-
\lambda{\boldsymbol{\sigma}},\mathbf{r}+(1-\lambda)
{\boldsymbol {\sigma}},\mathbf{v}_{1},\mathbf{v}_{2},t\right]
\nonumber\\
&- &\Delta \frac{m \sigma^d}{4}
\int \dd\mathbf{v}_{1}\int \dd\mathbf{v}_{2}\int
\dd\widehat{\boldsymbol {\sigma}}\Theta (\widehat{\boldsymbol{\sigma}}\cdot
\mathbf{g})(\widehat{\boldsymbol {\sigma}}\cdot \mathbf{g})\widehat{\boldsymbol {\sigma}}\nonumber\\
& \times& \left[\Delta +\al (\widehat{\boldsymbol {\sigma}}\cdot \mathbf{g})-2 (\widehat{\boldsymbol {\sigma}}\cdot \mathbf{G})\right]\nonumber\\
&\times&
\int_{0}^{1}\dd \lambda f_2\left[\mathbf{r}-
\lambda {\boldsymbol{\sigma}},\mathbf{r}+(1-\lambda)
{\boldsymbol {\sigma}},\mathbf{v}_{1},\mathbf{v}_{2},t\right].
\eeqa
Here, ${\bf G}=\frac{1}{2}({\bf V}_1+{\bf V}_2)$ is the velocity of the center of mass and $f_2$ is defined as
\begin{equation}
\label{1.16}
f_2({\bf r}_1, {\bf r}_2, {\bf v}_1, {\bf v}_2, t)\equiv \chi({\bf r}_1, {\bf r}_2)f({\bf r}_1, {\bf v}_1, t) f({\bf r}_2, {\bf v}_2, t).
\end{equation}
Finally, the cooling rate is given by
\beqa
\label{1.17}
\zeta&=&-\frac{m}{d n T}\sigma^{d-1}
\int \dd\mathbf{v}
_{1}\int \dd\mathbf{v}_{2}\int \dd\widehat{\boldsymbol {\sigma}}
\Theta (\widehat{\boldsymbol {\sigma}}\cdot
\mathbf{g})(\widehat{ \boldsymbol {\sigma}}\cdot
\mathbf{g})\nonumber\\
&\times&\left[\Delta^2+\al \Delta (\widehat{{\boldsymbol {\sigma}}}\cdot \mathbf{g})-\frac{1-\al^2}{4}(\widehat{{\boldsymbol {\sigma}}}\cdot \mathbf{g})^2\right]\nonumber\\
& \times &f_2(\mathbf{r}, \mathbf{r}+\boldsymbol {\sigma
},\mathbf{v}_{1},\mathbf{v}_{2},t).
\eeqa

The macroscopic balance equations \eqref{1.10}--\eqref{1.12} provide the basis for developing a hydrodynamic description of the confined granular fluid. Since those equations are not a closed set of equations for the hydrodynamic fields, one has to express $\mathsf{P}$, $\mathbf{q}$, and $\zeta$ as explicit functionals of the hydrodynamic  fields $n$, $\mathbf{U}$, and $T$ and their spatial gradients. This task can be accomplished by solving the Enskog equation \eqref{1.1} by means of the Chapman--Enskog method \cite{CC70} conveniently adapted to account for the dissipative dynamics.

\section{Chapman--Enskog method. First-order distribution function}
\label{sec2}

The Chapman--Enskog method assumes the existence of a \emph{normal} or hydrodynamic solution where all space and time dependence of the one-particle distribution function $f(\mathbf{r}, \mathbf{v}, t)$ only occurs through its functional dependence on the hydrodynamic fields. This means that
\beq
\label{2.1}
f(\mathbf{r}, \mathbf{v}, t)=f[\mathbf{v}|n(\mathbf{r},t), \mathbf{U}(\mathbf{r},t), T(\mathbf{r},t)].
\eeq
Note that, although energy is not conserved and the temperature is not strictly a slow field in the Delta-model, it has been shown in Ref.\ \cite{BBMG15} that after a short transient the distribution function does adopt a normal solution. A similar behavior is expected here for dense granular fluids.
The functional dependence \eqref{2.1} can be made local in space by means of an expansion in spatial gradients of the hydrodynamic fields. To
generate it, $f$ is written as a series expansion in a formal
parameter $\epsilon$ measuring the nonuniformity of the system,
\beq
\label{2.2}
f=f^{(0)}+\epsilon f^{(1)}+\epsilon^2 f^{(2)}+\cdots,
\eeq
where each factor of $\epsilon$ means an implicit gradient of a
hydrodynamic field. The uniformity parameter $\epsilon$ is related to
the Knudsen number defined by the length scale for variation
of the hydrodynamic fields.
Under some conditions, for the IHS and other undriven granular models, there is an intrinsic relation between collisional dissipation and some hydrodynamic gradients (e.g., in steady states such as the simple shear flow \cite{G03,SGD04}), which limits the application of the Chapman--Enskog expansion to regimes with small gradients (low Knudsen number). However,  homogeneous states are stable in the Delta-model for any inelasticity \cite{BRS13,BBGM16} and, hence, the strength of the gradients can be controlled by the initial or the boundary conditions as it happens for elastic media.  Thus, although our results will apply to sufficiently small gradients (low Knudsen number), they will not be restricted \emph{a priori} to small degree of dissipation.

According to the expansion \eqref{2.2} for the distribution
function, the Enskog collision operator and time derivative
are also expanded in powers of $\epsilon$:
\beq
\label{2.3}
J_\text{E}=J_\text{E}^{(0)}+\epsilon J_\text{E}^{(1)}+\cdots, \quad
\partial_t=\partial_t^{(0)}+\epsilon\partial_t^{(1)}+\cdots.
\eeq
The coefficients in the time derivative expansion are identified
by a representation of the fluxes and the cooling rate in the
macroscopic balance equations as a similar series through their
definitions as functionals of $f$. The expansion \eqref{2.2} yields
similar expansions for the momentum and heat fluxes, and the cooling rate when
substituted into their definitions \eqref{1.13}--\eqref{1.15} and \eqref{1.17}, respectively,
\begin{gather}
\label{2.4}
P_{ij}=P_{ij}^{(0)}+\epsilon P_{ij}^{(1)}+\cdots, \quad
\mathbf{q}=\mathbf{q}^{(0)}+\epsilon \mathbf{q}^{(1)}+\cdots,\\
\label{2.5}
\zeta=\zeta^{(0)}+\epsilon \zeta^{(1)}+\cdots.
\end{gather}
In this paper we shall restrict our calculations to the first order
in the uniformity parameter, which gives the Navier--Stokes transport coefficients.

\subsection{Zeroth-order approximation}

To zeroth-order in the expansion, $f^{(0)}$ verifies the kinetic equation
\beq
\label{2.6}
\partial_t^{(0)}f^{(0)}=J_\text{E}^{(0)}[f^{(0)},f^{(0)}],
\eeq
where
\beqa
\label{2.7}
& & J_\text{E}^{(0)}[f^{(0)},f^{(0)}]=\sigma^{d-1}\chi \int\dd{\bf v}_{2}\int \dd \widehat{\boldsymbol{\sigma}}
\Theta (-\widehat{{\boldsymbol {\sigma }}}\cdot {\bf g}-2\Delta)\nonumber\\
& &\times
(-\widehat{\boldsymbol {\sigma }}\cdot {\bf g}-2\Delta)
\al^{-2} f^{(0)}(\mathbf{v}_1'')f^{(0)}(\mathbf{v}_2'')-\sigma^{d-1}\chi\nonumber\\
& &\times
 \int\ \dd{\bf v}_{2}\int \dd\widehat{\boldsymbol{\sigma}}
\Theta (\widehat{{\boldsymbol {\sigma }}}\cdot {\bf g})
(\widehat{\boldsymbol {\sigma }}\cdot {\bf g})
f^{(0)}(\mathbf{v}_1)f^{(0)}(\mathbf{v}_2).
\eeqa
Here, $\chi$ refers to the the pair correlation evaluated
with all density fields at the local point $\mathbf{r}$. The collision
operator \eqref{2.7} can be recognized as the Boltzmann collision operator
for the collisional model multiplied by this factor $\chi$. The macroscopic
balance equations to this order read
\beq
\label{2.8}
\partial_t^{(0)}n=\partial_t^{(0)}U_i=0, \quad \partial_t^{(0)}T=-T\zeta^{(0)}.
\eeq
Since $f^{(0)}$ qualifies as a normal solution, then
\beq
\label{2.10}
\partial_t^{(0)}f^{(0)}=-\zeta^{(0)}T \partial_T f^{(0)}
\eeq
and Eq.\ \eqref{2.6} reads
\beq
\label{2.11}
-\zeta^{(0)}T \partial_T f^{(0)}=J_\text{E}^{(0)}[f^{(0)},f^{(0)}].
\eeq
The solution to Eq.\ \eqref{2.10} has been widely studied in Refs.\ \cite{BGMB13,BMGB14,BGMB14} where it
has been shown that it  adopts the scaled form
\beq
\label{2.12a}
f^{(0)}(\mathbf{r}, \mathbf{v},t)=n(\mathbf{r},t) v_0(\mathbf{r},t)^{-d}
\varphi(\mathbf{c},\Delta^*),
\eeq
where $v_0=\sqrt{2T/m}$, $\mathbf{c}\equiv \mathbf{v}/v_0$, and $\Delta^* \equiv \Delta/v_0$. Thus, in contrast to the freely cooling IHS model, the unknown scaled distribution $\varphi$ depends on the granular temperature $T$ not only through the scaled velocity $\mathbf{c}$ but also through the dimensionless parameter $\Delta^*(t) \propto T(t)^{-1/2}$.  Then
\beq
\label{2.12}
T\partial_T  f^{(0)}=-\frac{1}{2}\frac{\partial }{\partial \mathbf{V}}\cdot \left(\mathbf{V}f^{(0)}\right)
-\frac{1}{2}\Delta^*\frac{\partial f^{(0)}}{\partial \Delta^*}.
\eeq

An exact solution to Eq.\ \eqref{2.11} has not been found so far. However, a
very good approximation can be obtained from an expansion
in Sonine polynomials. In particular, the kurtosis
\beq
\label{2.13}
a_2=\frac{4}{d(d+2)}\int\; \dd \mathbf{c}\; c^4 \varphi(c)-1
\eeq
of the scaled distribution $\varphi$ has been estimated in Ref.\ \cite{BGMB13}. In all of the following it is presumed that the distribution $f^{(0)}$ is known. Since the distribution function is isotropic, the zeroth order pressure tensor and heat flux are found from Eqs.\ \eqref{1.13}--\eqref{1.15} to be
\beq
\label{2.14}
P_{ij}^{(0)}=p\delta_{ij}, \quad \mathbf{q}^{(0)}=\mathbf{0},
\eeq
where the hydrostatic pressure can be written as $p=n T p^*$ where
\beqa
\label{2.15}
p^*&=&1+2^{d-2}\chi \phi (1+\al)+\frac{2^{d-1}\Gamma\left(\frac{d}{2}\right)}{\sqrt{\pi}
\Gamma\left(\frac{d+1}{2}\right)}\chi \phi \Delta^*
\nonumber\\
&\times& \int \dd\mathbf{c}_1
\int \dd\mathbf{c}_2  g^*
\varphi(\mathbf{c}_1)\varphi(\mathbf{c}_2).
\eeqa
Here,
\beq
\label{2.16}
\phi=\frac{\pi^{d/2}}{2^{d-1}d \Gamma(d/2)} n\sigma^d
\eeq
is the solid volume fraction and $\mathbf{g}^*\equiv \mathbf{g}/v_0$.
Note that, besides the standard ideal gas and excluded volume contributions to the pressure, there is a new term proportional to $\Delta$, which is due to the additional momentum transfer at collisions.

Finally, the zeroth-order contribution to the cooling rate is
\beqa
\label{2.17}
\zeta^{(0)}&=&-\frac{2}{d}n\sigma^{d-1}v_0 \chi \int \dd\mathbf{c}_1
\int \dd\mathbf{c}_2 \varphi(\mathbf{c}_1)\varphi(\mathbf{c}_2)\nonumber\\
& & \times \left(B_1 g^{*}\Delta^{*2}+B_2 \al g^{*2}\Delta^*-\frac{1-\al^2}{4}B_3 g^{*3}\right),
\nonumber\\
\eeqa
where we have introduced the quantities \cite{NE98}
\beq
\label{2.18}
B_k\equiv \int\; \dd\widehat{\boldsymbol{\sigma}}\,
\Theta (\widehat{{\boldsymbol {\sigma }}}\cdot \mathbf{g})(\widehat{\boldsymbol {\sigma }}\cdot \widehat{\mathbf{g}})^k=\pi^{(d-1)/2} \frac{\Gamma\left(\frac{k+1}{2}\right)}{\Gamma\left(\frac{k+d}{2}\right)}
\eeq
for positive integers $k$.

\subsection{First-order approximation}

The first-order distribution $f^{(1)}$ can be obtained by following similar steps as those made before for the freely cooling IHS model. The main new feature of the first-order solution is that there are new terms coming from the additional time-dependence of $f^{(0)}$ through $\Delta^*$. The first-order velocity distribution function $f^{(1)}$ is given by
\begin{eqnarray}
\label{2.19}
f^{(1)}&=&\boldsymbol{\mathcal{A}}\left(
\mathbf{V}\right)\cdot  \nabla \ln
T+\boldsymbol{\mathcal{B}}\left(
\mathbf{V}\right) \cdot \nabla \ln n
\nonumber\\
& & +\mathcal{C}_{ij}\left( \mathbf{V} \right)\frac{1}{2}\left( \partial _{i}U_{j}+\partial _{j
}U_{i}-\frac{2}{d}\delta _{ij}\nabla \cdot
\mathbf{U} \right)\nonumber\\
& & +\mathcal{D}\left( \mathbf{V} \right) \nabla \cdot\mathbf{U}.
\end{eqnarray}
The quantities $\boldsymbol{\mathcal{A}}$, $\boldsymbol{\mathcal{B}}$, $\mathcal{C}_{ij}$ and $\mathcal{D}$ are the solutions of the following linear integral equations:
\begin{gather}
-\zeta^{(0)}T\frac{\partial \boldsymbol{\mathcal{A}}}{\partial T}-\boldsymbol{\mathcal{A}}T\frac{\partial \zeta^{(0)}}
{\partial T} +
\mathcal{L}\boldsymbol{\mathcal{A}}=\mathbf{A},
\label{2.20}\\
-\zeta^{(0)}T\frac{\partial \boldsymbol{\mathcal{B}}}{\partial T}+
\mathcal{L}\boldsymbol{\mathcal{B}}=\mathbf{B}+\zeta^{(0)}\left(1+\phi\frac{\partial}{\partial \phi}\ln \chi\right)\boldsymbol{\mathcal{A}},  \label{2.21}\\
-\zeta^{(0)}T\frac{\partial \mathcal{C}_{ij}}{\partial T}
+\mathcal{L}\mathcal{C}_{ij}=C_{ij},  \label{2.22}\\
-\zeta^{(0)}T\frac{\partial \mathcal{D}}{\partial T}+
\mathcal{L}\mathcal{D}=D.  \label{2.23}
\end{gather}
In Eqs.\ \eqref{2.20}--\eqref{2.23}, the linear operator $\mathcal{L}$ is given by
\begin{equation}
\mathcal{L}X=-\left(J_\text{E}^{(0)}[f^{(0)},X]+J_\text{E}^{(0)}[X,f^{(0)}]\right),  \label{2.24}
\end{equation}
while the inhomogeneous terms, which depend on $f^{(0)}$, are defined by
\beqa
{\bf A}\left( \mathbf{V}\right)&=&-\mathbf{V}T\frac{\partial f^{(0)}}{\partial T}
-\frac{p}{\rho}\left(1+T\frac{\partial}{ \partial T}\ln p^*\right)\frac{\partial f^{(0)}}{\partial \mathbf{V}}\nonumber\\
& &-\boldsymbol{\mathcal{K}}\left[T\frac{\partial f^{(0)}}{\partial T}\right],  \label{2.25}
\eeqa
\beqa
{\bf B}\left(\mathbf{V}\right)&=& -{\bf V}f^{(0)}-\frac{p}{\rho}
\left(1+\phi\frac{\partial}{\partial \phi}\ln p^*\right)
\frac{\partial f^{(0)}}{\partial \mathbf{V}}\nonumber\\
& & 
-\left(1+\frac{1}{2}\phi\frac{\partial}{\partial \phi}\ln \chi\right)
\boldsymbol{\mathcal{K}}\left[f^{(0)}\right], 
\label{2.26}
\eeqa
\beq
\label{2.27}
C_{ij}\left(\mathbf{V}\right)=V_i\frac{\partial f^{(0)}}{\partial V_j}+\mathcal{K}_i\left[\frac{\partial f^{(0)}}{\partial V_j}
\right],
\eeq
\beqa
D\left(\mathbf{V}\right)&=&\frac{1}{d}\frac{\partial}{\partial \mathbf{V}}\cdot \left( \mathbf{V} f^{(0)}\right)
+\left(\zeta_{U}+\frac{2}{d} p^*\right)T\frac{\partial f^{(0)}}{\partial T} \nonumber\\
& &+\frac{1}{d}\mathcal{K}_{i}\left[\frac{\partial f^{(0)}}{\partial V_i}\right].   \label{2.28}
\eeqa
The operator $\mathcal{K}_{i}$ is
\beqa
\mathcal{K}_{i}[X] &=&-\sigma^{d}\chi\int \dd \mathbf{v}_{2}\int \dd\widehat{\boldsymbol {\sigma
}}\Theta (-\widehat{\boldsymbol {\sigma}} \cdot
\mathbf{g}-2\Delta)\nonumber\\
& &\times (-\widehat{\boldsymbol {\sigma }}\cdot
\mathbf{g}-2\Delta)
\widehat{\sigma}_i \alpha^{-2}f^{(0)}(\mathbf{v}_{1}'')X(\mathbf{v}_{2}'')\nonumber\\
& &+\sigma^{d}\chi\int \dd \mathbf{v}_{2}\int \dd\widehat{\boldsymbol {\sigma
}}\Theta (\widehat{\boldsymbol {\sigma}} \cdot
\mathbf{g})(\widehat{\boldsymbol {\sigma }}\cdot
\mathbf{g})
\widehat{\sigma}_i \nonumber\\
& & \times f^{(0)}(\mathbf{v}_{1})X(\mathbf{v}_{2}).  \label{2.29}
\nonumber\\
\eeqa
In Eq.\ \eqref{2.28}, $\zeta_U$ is defined through the expression
\begin{equation}
\label{2.15b}
\zeta=\zeta^{(0)}+\zeta_U\nabla \cdot {\bf U},
\end{equation}
where $\zeta^{(0)}$ is given by Eq.\ \eqref{2.17}.

In the low-density limit ($\phi=0$), $p^*=1$, ${\cal K}_i \to 0$, and the integral equations \eqref{2.20}--\eqref{2.23} are consistent with those obtained in Ref.\ \cite{BBMG15}. In addition, even for dilute granular gases there is a first-order contribution to the cooling rate. This is because in this limit ($\phi\to 0$) the quantity $D$ becomes
\beq
\label{2.17b}
D=\zeta_U T\frac{\partial f^{(0)}}{\partial T}-\frac{1}{d}\Delta^*\frac{\partial f^{(0)}}{\partial \Delta^*},
\eeq
and hence, the integral equation \eqref{2.23} has a nonzero solution. Note that $\zeta_U \neq 0$ when $\phi \neq 0$ for the IHS model \cite{GD99,L05}.

The next step is to obtain the explicit expressions of the Navier--Stokes transport coefficients. These coefficients are given in terms of the solutions of the linear integral equations \eqref{2.20}--\eqref{2.23}.

\section{Navier--Stokes transport coefficients}
\label{sec3}

\subsection{Pressure tensor}

To first order in the spatial gradients, the pressure tensor is given by
\begin{equation}
\label{3.1}
P_{ij}^{(1)}=-\eta\left( \partial _{i}U_{j}+\partial _{j}U_{i}-\frac{2}{d}\delta _{ij}\nabla \cdot
\mathbf{U} \right) -\gamma  \nabla \cdot \mathbf{U} \delta_{ij},
\end{equation}
where $\eta$ is the shear viscosity and $\gamma$ is the bulk viscosity. These coefficients have kinetic and collisional contributions, i.e., $\eta=\eta_\text{k}+\eta_\text{c}$ and $\gamma=\gamma_\text{c}$ since $\gamma_\text{k}=0$.
The collisional contributions (obtained in Appendix~\ref{appA}) are given by
\beq
\label{3.2}
\eta_\text{c}=n\sigma^d \chi \left[\frac{B_2}{d+2}(1+\al)+\frac{B_1}{d+1}\Delta^*I_\eta
\right]\eta_\text{k}+\frac{d}{d+2}\gamma,
\eeq
\beq
\label{3.3}
\gamma=n^2\sigma^{d+1}m \chi v_0\left[B_3\frac{d+1}{4d^2}(1+\al)
I_\gamma+\frac{B_2}{2d}\Delta^*\right],
\eeq
where the dimensionless integrals in Eqs.\ \eqref{3.2} and \eqref{3.3}, after applying the approximation \eqref{a4}, are
\begin{align}
\label{3.3.1}
I_\eta&=\int \dd\mathbf{c}_1
\int \dd\mathbf{c}_2\; g^{*-1}g_x^{*2}g_y^{*2}\varphi_\text{M}(\mathbf{c}_1)\varphi_\text{M}(\mathbf{c}_2),\\
\label{3.3.2}
I_\gamma&=\int \dd \mathbf{c}_1
\int \dd\mathbf{c}_2\; g^*\; \varphi(\mathbf{c}_1)\varphi(\mathbf{c}_2),
\end{align}
with $\varphi_\text{M}(\mathbf{c})=\pi^{-d/2}e^{-c^2}$.

The kinetic contribution  $\eta_\text{k}$ to the shear viscosity is defined as
\beq
\label{3.4}
\eta_\text{k}=- \frac{1}{(d-1)(d+2)}\int\; \dd \mathbf{v}\; D_{ij}(\mathbf{V})\; \mathcal{C}_{ij}(\mathbf{V}),
\eeq
where $D_{ij}(\mathbf{V})$ is the traceless tensor
\beq
\label{3.5}
D_{ij}(\mathbf{V})=m\left(V_iV_j-\frac{1}{d}\delta_{ij}V^2\right).
\eeq
The expression of $\eta_\text{k}$ can be obtained by multiplying both sides of Eq.\ \eqref{2.22} by $D_{ij}$ and integrating over velocity. The result is
\beq
\label{3.6}
\left(-\zeta^{(0)}T\partial_T+\nu_\eta\right)\eta_\text{k}=-\frac{\int \dd\mathbf{V}\; D_{ij}(\mathbf{V}) C_{ij}(\mathbf{V})}{(d-1)(d+2)},
\eeq
where
\begin{equation}
\label{3.7}
\nu_\eta=\frac{\int \dd{\bf v} D_{ij}({\bf V}){\cal L}{\cal C}_{ij}({\bf V})}
{\int \dd{\bf v}D_{ij}({\bf V}){\cal C}_{ij}({\bf V})}.
\end{equation}
The kinetic coefficient $\eta_\text{k}$ can be written as $\eta_\text{k}(T)=\eta_0(T)\eta_\text{k}^*(\Delta^*)$, where
\begin{equation}
\label{3.8}
\eta_0(T)=\frac{d+2}{8}\Gamma\left(\frac{d}{2}\right)
\pi^{-\frac{d-1}{2}}\sigma^{1-d}\sqrt{mT}
\end{equation}
is the low density value of the shear viscosity in the elastic limit. Thus,
\beq
\label{3.9}
T\partial_T \eta_\text{k}=\frac{1}{2}\eta_\text{k}-\frac{1}{2}\eta_\text{k} \Delta^*
\frac{\partial \ln \eta_\text{k}^*}{\partial \Delta^*},
\eeq
and Eq.\ \eqref{3.6} reads
\begin{multline}
\label{3.10}
\frac{1}{2}\zeta^{(0)}\eta_\text{k} \Delta^* \frac{\partial \ln \eta_\text{k}^*}{\partial \Delta^*}+\left(\nu_\eta-\frac{1}{2}\zeta^{(0)}\right)\eta_\text{k}=n T\\
 -\frac{1}{(d-1)(d+2)}\int\; \dd \mathbf{v}
D_{ij}(\mathbf{V}) {\cal K}_i\left[\frac{\partial f^{(0)}}{\partial V_j}
\right],
\end{multline}
where use has been made of the explicit form of $C_{ij}$. As expected \cite{BBMG15}, in contrast to the conventional IHS model, $\eta_\text{k}$ is given as the solution of an intricate first-order differential equation. The integral appearing in the right-hand side of Eq.\ \eqref{3.10} has been computed in  Appendix \ref{appB} with the result
\begin{multline}
\label{3.8b}
\int\; \dd \mathbf{v}
D_{ij}(\mathbf{V}) {\cal K}_i\left[\frac{\partial f^{(0)}}{\partial V_j}\right]=\\
2^{d-2}(d-1)\chi \phi (1+\al)(1-3\al)n T\\
+2^d (d-1)\chi \phi \Delta^* n T\left[\frac{\Gamma\left(\frac{d}{2}\right)}{\sqrt{\pi}
\Gamma\left(\frac{d+1}{2}\right)} {I}_\eta' - \Delta^{*}\right],
\end{multline}
where
\beq
\label{3.9b}
{I}_\eta'=\int \dd{\bf c}_1\int \dd{\bf c}_2
\varphi({\bf c}_1)\varphi({\bf c}_2)
\left[2g^{*-1} (\mathbf{g}^*\cdot \mathbf{c}_1)-2(1+\al)g^*\right].
\eeq

\subsection{Heat flux}

The constitutive form for the heat flux in the Navier--Stokes approximation is
\begin{equation}
\label{3.10b}
{\bf q}^{(1)}=-\kappa \nabla T-\mu \nabla n,
\end{equation}
where $\kappa$ is the thermal conductivity and $\mu$ is the diffusive heat conductivity coefficient. The coefficient $\mu$ is an additional transport coefficient not present in the elastic case. Both transport coefficients $\kappa$ and $\mu$ have kinetic and collisional contributions.

The collisional contributions $\kappa_\text{c}$ and $\mu_\text{c}$ have been determined in Appendix~\ref{appA}. They can be written as
\begin{align}
\kappa_\text{c}=& n\sigma^d \chi  \left[\frac{3}{2}\frac{B_2}{d+2}(1+\al)+
\frac{8B_1}{d(d+1)(d+2)}\Delta^* I_\kappa'\right]\kappa_\text{k}\nonumber\\
&+\frac{m\sigma^{d+1}}{8dT}\chi n^2 v_0^3\left[B_3(1+\al) I_\kappa+2B_2 \Delta^* I_\kappa''\right],\label{3.11}\\
\mu_\text{c}=& n\sigma^d \chi  \left[\frac{3}{2}\frac{B_2}{d+2}(1+\al)+
\frac{8B_1}{d(d+1)(d+2)}\Delta^* I_\kappa'\right]\mu_\text{k}, \label{3.12}
\end{align}
where, after applying the approximations \eqref{a9}, the dimensionless integrals $I_\kappa$, $I_\kappa'$, and $I_\kappa''$ are given by
\begin{widetext}
\begin{align}
\label{3.13}
I_\kappa&=\int\ \dd\mathbf{c}_1 \int\dd\mathbf{c}_2\; \varphi(\mathbf{c}_1)
\varphi(\mathbf{c}_2)\left[g^{*-1}({\bf g}^*\cdot {\bf G}^*)^{2}+g^*G^{*2}+\frac{3}{2}g^*({\bf g}^*\cdot {\bf G}^*)+\frac{1}{4}g^{*3}\right],\\
\label{3.14}
I_\kappa'&=\int \dd\mathbf{c}_1 \int\dd\mathbf{c}_2 \; \varphi_\text{M}(\mathbf{c}_1)\varphi(\mathbf{c}_2)g^{*-1}\left[
(\mathbf{g}^*\cdot \mathbf{S}^*)(\mathbf{g}^*\cdot \mathbf{G}^*)+g^{*2}(\mathbf{G}^*\cdot \mathbf{S}^*)\right],\\
\label{3.15}
I_\kappa''&=\frac{d}{2}+\Delta^*\;
\int \dd\mathbf{c}_1 \int\dd\mathbf{c}_2 \; \varphi(\mathbf{c}_1)\;\frac{\partial \varphi(\mathbf{c}_2)}{\partial \Delta^*}\;(\mathbf{g}^*\cdot \mathbf{G}^*),
\end{align}
\end{widetext}
where $\mathbf{G}^*\equiv \mathbf{G}/v_0$ and
\begin{equation}
\label{3.16}
\mathbf{S}^*(\mathbf{c}_1)=\left(c_1^2-\frac{d+2}{2}\right){\bf c}_1.
\end{equation}

The kinetic parts $\kappa_\text{k}$ and $\mu_\text{k}$ are defined as
\begin{align}
\label{3.17}
\kappa_\text{k}&=-\frac{1}{dT}\int\, \dd{\bf v} {\bf S}({\bf V})\cdot {\boldsymbol {\cal A}}({\bf V}),\\
\label{3.18}
\mu_\text{k}&=-\frac{1}{dn}\int\, \dd{\bf v} {\bf S}({\bf V})\cdot {\boldsymbol {\cal B}}({\bf V}),
\end{align}
where
\begin{equation}
\label{3.19}
{\bf S}({\bf V})=\left(\frac{m}{2}V^2-\frac{d+2}{2}T\right){\bf V}.
\end{equation}

The kinetic part of the thermal conductivity is obtained by multiplication of Eq.\ (\ref{2.25}) by ${\bf S}({\bf V})$ and integration over the velocity. The result is
\beqa
\label{3.20}
& &\frac{1}{2}\zeta^{(0)}\kappa_\text{k} \Delta^* \frac{\partial \ln \kappa_\text{k}^*}{\partial \Delta^*}+\left(\nu_\kappa+
\frac{1}{2}\zeta^{(0)} \Delta^* \frac{\partial \ln \zeta_0^*}{\partial \Delta^*}\right.\nonumber\\
& &\left.-2\zeta^{(0)}\right)\kappa_\text{k}=
-\frac{1}{d T}\int\; \dd \mathbf{v}\mathbf{S}(\mathbf{V})\cdot \mathbf{A}(\mathbf{V}),
\eeqa
where $\zeta_0^*\equiv \zeta^{(0)}/\nu_0$, $\nu_0=n T/\eta_0$, $\kappa_\text{k}^*\equiv \kappa_\text{k}/\kappa_0$ and
\begin{equation}
\label{3.21}
\nu_\kappa=\frac{\int \dd{\bf v} {\bf S}({\bf V})\cdot {\cal L}\boldsymbol{\mathcal{A}}({\bf V})}
{\int \dd{\bf v}{\bf S}({\bf V})\cdot \boldsymbol{\mathcal{A}}({\bf V})}.
\end{equation}
Here,
\begin{equation}
\label{3.22}
\kappa_0=\frac{d(d+2)}{2(d-1)}\frac{\eta_0}{m}
\end{equation}
is the low density value of the thermal conductivity of an elastic gas. The right-hand side of Eq.\ \eqref{3.20} can be written more explicitly as
\begin{widetext}
\beqa
\label{3.23}
-\frac{1}{d T}\int\; \dd \mathbf{v}\mathbf{S}(\mathbf{V})\cdot \mathbf{A}(\mathbf{V})&=&\frac{d+2}{2m}n T\left(1+2 a_2-\frac{1}{2}\Delta^* \frac{\partial a_2}{\partial \Delta^*}\right)-\frac{1}{2 d T}
\int\; \dd{\bf v}{\bf S}({\bf V})\cdot \boldsymbol{\mathcal{K}}\left[\frac{\partial}{\partial \mathbf{V}}\cdot \left( \mathbf{V}f^{(0)}\right)\right]\nonumber\\
& &-\frac{1}{2 d T}\Delta^*\frac{\partial}{\partial \Delta^*}
\int\; \dd{\bf v}{\bf S}({\bf V})\cdot \boldsymbol{\mathcal{K}}\left[f^{(0)}\right],
\eeqa
where $a_2$ is defined by Eq.\ \eqref{2.13}. The last two integrals of the right hand side of Eq.\ \eqref{3.23} have been evaluated in Appendix~\ref{appB} by assuming $a_2=0$ for the sake of simplicity.

To determine $\mu_\text{k}$, Eq.\ \eqref{2.26} is multiplied by ${\bf S}({\bf V})$ and integrated over the velocity to get
\beqa
\label{3.24}
\frac{1}{2}\zeta^{(0)}\mu_\text{k} \Delta^* \frac{\partial \ln \mu_\text{k}^*}{\partial \Delta^*}+\left(\nu_\mu-
\frac{3}{2}\zeta^{(0)}\right)\mu_\text{k}&=&\frac{d+2}{2}\frac{T^2}{m}a_2
+\frac{T\zeta^{(0)}}{n}\left(1+\phi\partial_\phi\ln \chi\right)
\kappa_\text{k}+\left(1+\frac{1}{2}\phi\partial_\phi\ln \chi\right)
\nonumber\\
& & \times \frac{1}{dn}\int\; \dd \mathbf{v}\mathbf{S}(\mathbf{V})\cdot \boldsymbol{\mathcal{K}}\left[f^{(0)}\right],
\eeqa
\end{widetext}
where $\mu_\text{k}^*\equiv (n/T \kappa_0)\mu_\text{k}$ and
\begin{equation}
\label{3.25}
\nu_\mu=\frac{\int \dd{\bf v} {\bf S}({\bf V})\cdot {\cal L}\boldsymbol{\mathcal{B}}({\bf V})}
{\int \dd{\bf v}{\bf S}({\bf V})\cdot \boldsymbol{\mathcal{B}}({\bf V})}.
\end{equation}

In summary, the results obtained in this section provide the collisional transfer contributions to the Navier--Stokes transport coefficients in terms of integrals involving the scaled distribution $\varphi$ [see Eqs.\ \eqref{3.3.1}, \eqref{3.3.2}, \eqref{3.9b}, \eqref{3.13}, \eqref{3.14}, and \eqref{3.16}] while their kinetic contributions are given in terms of the numerical solutions of the differential equations \eqref{3.9}, \eqref{3.20}, and \eqref{3.24}. A detailed study of the dependence of $\eta_\text{k}^*$, $\kappa_\text{k}^*$, and $\mu_\text{k}^*$ on both $\alpha$ and $\Delta^*$ has been carried out in Ref.\ \cite{BBMG15} for a low-density gas.


\section{Transport coefficients at the stationary temperature. Two-dimensional case}
\label{sec4}

As mentioned before, the explicit dependence of the (reduced) Navier-Stokes transport coefficients on the solid volume fraction and the coefficient of restitution requires knowledge of the quantities $\zeta^{(0)}$, $a_2$, $\nu_\eta$, $\nu_\kappa$, and $\nu_\mu$, the dimensionless integrals $\left\{I_\eta, I_\gamma, I_\eta', I_\kappa, I_\kappa', I_\kappa''\right\}$, and two integrals involving the operator $\boldsymbol{\mathcal{K}}$. All these quantities are given in terms of the zeroth-order scaled distribution $\varphi$ and the solutions $\boldsymbol{\mathcal{A}}$, $\boldsymbol{\mathcal{B}}$, and $\mathcal{C}_{ij}$ to the linear integral equations \eqref{3.10}, \eqref{3.20}, and \eqref{3.24}.

\begin{figure}[h]
{\includegraphics[width=0.85\columnwidth]{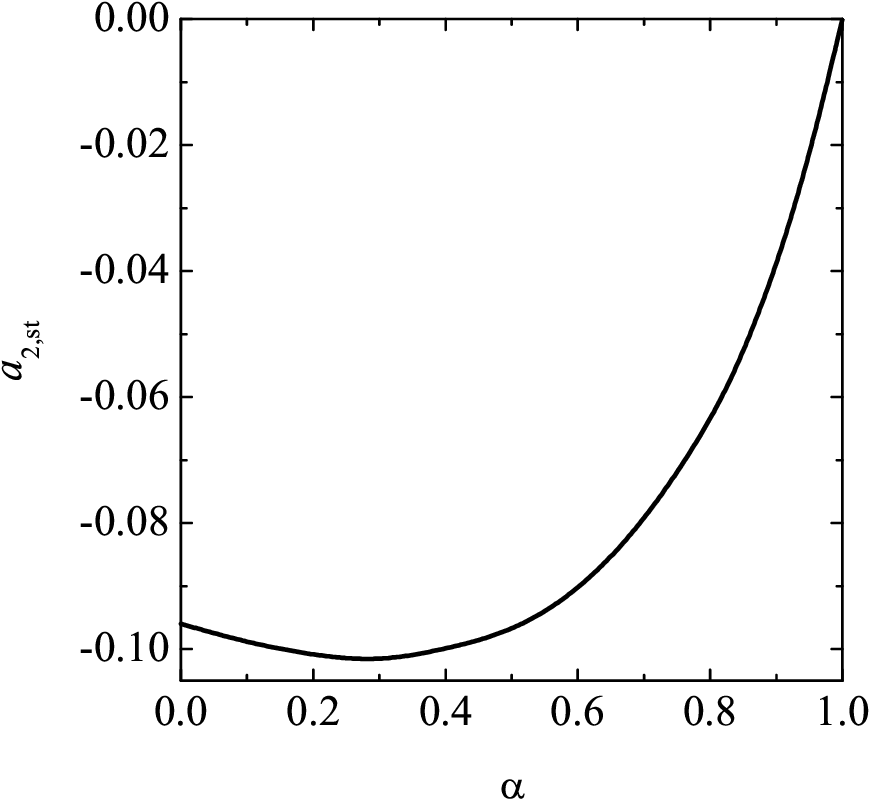}}
\caption{Dependence of the Sonine coefficient $a_{2,\text{s}}$ on the coefficient of restitution $\al$ for a two-dimensional ($d=2$) granular gas at the stationary temperature.}
\label{a2fig}
\end{figure}

Determination of $\varphi$ was discussed in Refs.\ \cite{BMGB14,BBMG15}. A good approximation to the zeroth-order solution is provided by the so-called first Sonine approximation
\beqa
\label{4.0}
\varphi(\mathbf{c},\Delta^*)&\approx& \varphi_\text{M}(\mathbf{c})\left\{1+a_2(\Delta^*)\left[\frac{c^4}{2}-\frac{d+2}{2}c^2\right.\right.\nonumber\\
& & \left.\left.+\frac{d(d+2)}{8}\right]\right\},
\eeqa
where the Sonine coefficient $a_2$ is defined in Eq.\ \eqref{2.13}. The dependence of the coefficient $a_2$ on $\Delta^*$ has been extensively studied in Ref.\ \cite{BMGB14}. When the quadratic terms in $a_2$ are neglected, it is shown that $a_2$ obeys a differential equation where the coefficients of this equation are nonlinear functions of both $\alpha$ and $\Delta^*$. This equation has been numerically solved for different initial conditions in order to identify the common hydrodynamic solution (see Fig.\ 1 of Ref.\ \cite{BMGB14}). In particular, the evaluation of $a_2$ can be performed in a quite accurate way in the relevant state with stationary temperature. In the vicinity of the steady state, $|\partial a_2/\partial \Delta^*|\ll 1$ and hence an explicit expression of $a_2$ can be displayed \cite{BMGB14}. The explicit dependence of $a_2$ on $\al$ in the steady state can be easily derived by imposing locally $\partial_t^{(0)}T=0$. According to the last identity in Eq.\ \eqref{2.8}, this implies that $\zeta^{(0)}=0$. The zeroth-order contribution to the cooling rate can be estimated from Eq.\ \eqref{2.17} by replacing $\varphi \to \varphi_\text{M}$  for the sake of simplicity. The result is
\beq
\label{4.8}
\zeta^{(0)}=\sqrt{\frac{\pi}{2}}n\sigma v_0 \chi \left(1-\al^2-2\Delta_\text{s}^{*2}-\sqrt{2\pi}\al \Delta_\text{s}^*\right),
\eeq
where $\Delta_\text{s}^*$ refers to the steady value of $\Delta^*$. The condition $\zeta^{(0)}=0$ yields a quadratic equation in $\Delta_\text{s}^*$, whose physical solution (i.e., $\Delta_\text{s}^*=0$ when $\al=1$) is
\beq
\label{4.9}
\Delta_\text{s}^*(\al)=\frac{1}{2}\sqrt{\frac{\pi}{2}}\al\left[\sqrt{1+
\frac{4(1-\al^2)}{\pi \al^2}}-1\right].
\eeq
Once the $\alpha$-dependence of $\Delta_\text{s}^*(\al)$ is known, the explicit form of $a_2$ near the steady state can be obtained \cite{BMGB14}. In particular, for a two-dimensional ($d=2$) granular gas, the kurtosis $a_{2,\text{s}}$ can be written as
\begin{widetext}
\beq
\label{4.0.1}
a_{2,\text{s}}=-16 \frac{2\al^4+3\sqrt{2\pi}\Delta_\text{s}^*\al^3+3(4\Delta_\text{s}^{*2}-1)\al^2+\sqrt{2\pi}\Delta_\text{s}^*
(4\Delta_\text{s}^{*2}-3)\al+2\Delta_\text{s}^{*2}(2\Delta_\text{s}^{*2}-3)+1}
{30\al^4+24\sqrt{2\pi}\Delta_\text{s}^*\al^3+(36\Delta_\text{s}^{*2}-5)\al^2-8(4+5\sqrt{2\pi}\Delta_\text{s}^*)\al-\Delta_\text{s}^*\left(
4\Delta_\text{s}^{*3}+162 \Delta_\text{s}^{*}+16\sqrt{2\pi}\right)-57},
\eeq
\end{widetext}
where $\Delta_\text{s}^{*}$ is given by Eq.\ \eqref{4.9}. Dependence of $a_{2,\text{s}}$ on $\al$ is plotted in Fig~\ref{a2fig} for $d=2$. As shown in the figure, the coefficient $a_{2,\text{s}}$ is always negative and its magnitude never exceeds 0.103. Thus, in the steady state, contributions to the transport coefficients coming from the terms proportional to $a_{2,\text{s}}$ are negligible as compared with the remaining contributions (see, however, Sect.\ \ref{sec5} for a further discussion). As a consequence, given that a theory incorporating these non-Gaussian corrections is not needed in practice for computing the Navier--Stokes transport coefficients of the confined granular gas, henceforth the integrals involving the scaled distribution $\varphi$ will be computed by replacing it by its Maxwellian form $\varphi_\text{M}$.

Although our theory applies for an arbitrary number of dimensions $d$, we are mainly interested in a two-dimensional confined system. For this reason all the results provided in this section will be restricted to $d=2$. In particular, the dimensionless integrals $I_\eta$, $I_\gamma$, ${I}_\eta'$, $I_\kappa$, ${I}_\kappa'$, and ${I}_\kappa''$ are given by
\begin{gather}
\label{4.1}
I_\eta=\frac{3}{8}\sqrt{\frac{\pi}{2}}, \quad I_\gamma=\sqrt{\frac{\pi}{2}},
\quad {I}_\eta'=-\sqrt{\frac{\pi}{2}}\left(1+2\al\right),\\
\label{4.2}
I_\kappa=\frac{3}{4}\sqrt{2\pi}, \quad I_\kappa'=\frac{9}{16}\sqrt{\frac{\pi}{2}}, \quad {I}_\kappa''=1.
\end{gather}

Finally, to evaluate the kinetic parts of the transport coefficients, one still needs to know the explicit forms of the collision frequencies $\nu_\eta$, $\nu_\kappa$ and $\nu_\mu$ as well as the integrals involving the operator $\boldsymbol{\mathcal{K}}[X]$. In the case of the collision frequencies, one takes the approximations \eqref{a4} for $\mathcal{C}_{ij}(\mathbf{V})$ and \eqref{a9} for $\boldsymbol{\mathcal{A}}$ and $\boldsymbol{\mathcal{B}}$. These integrals have been performed in Ref.\ \cite{BBMG15} for a $d$-dimensional system. In the case $d=2$, one gets
\beq
\label{4.3}
\nu_\eta^*=\frac{3}{8}\chi \left[\left(\frac{7}{3}-\alpha\right)(1+\alpha)
+\frac{2\sqrt{2\pi}}{3}(1-\al)\Delta^*-\frac{2}{3}\Delta^{*2}\right],
\eeq
\beqa
\label{4.4}
\nu_\kappa^*&=&\nu_\mu^*=\frac{1+\alpha}{2}\chi\left[
\frac{1}{2}+\frac{15}{8}(1-\alpha)\right]\nonumber\\
& & -\frac{\Delta^*}{16}\chi
\left[\sqrt{2\pi}(5\al-1)+10\Delta^{*}\right],
\eeqa
where $\nu_\eta^*\equiv \nu_\eta/\nu_0$, $\nu_\kappa^*\equiv \nu_\kappa/\nu_0$, and $\nu_\mu^*\equiv \nu_\mu/\nu_0$. Moreover, the expressions derived in Appendix~\ref{appB} for $d=2$ read
\begin{widetext}
\begin{align}
\label{4.5}
\int\dd \mathbf{V}
D_{ij}(\mathbf{V}) {\cal K}_i\left[\frac{\partial f^{(0)}}{\partial V_j}\right]
=&\phi \chi n T \left[(1+\al)(1-3\al)-4\sqrt{\frac{2}{\pi}}(1+2\al)\Delta^*-4\Delta^{*2}\right],\\
\label{4.6}
\int \dd{\bf V}{\bf S}({\bf V})\cdot \boldsymbol{\mathcal{K}}\left[\frac{\partial}{\partial \mathbf{V}}\cdot \left( \mathbf{V}f^{(0)}\right)\right]=&8\phi \chi \frac{n T^2}{m}\bigg\{\frac{3}{8}(1+\al)^2(1-2\al)+\frac{\Delta^*}{\sqrt{2\pi}}\bigg[\frac{3}{4}+3(1+\al)\left(1-\sqrt{\frac{\pi}{2}}
\Delta^*\right)\nonumber\\
 & -\frac{9}{2}(1+\al)^2-\Delta^{*2}\bigg]\bigg\},\\
\label{4.7}
\int\dd \mathbf{v}\mathbf{S}(\mathbf{V})\cdot \boldsymbol{\mathcal{K}}\left[f^{(0)}\right]=&-8\phi \chi \frac{n T^2}{m}\left\{\frac{3}{8}\al(1-\al^2)-\frac{\Delta^*}{\sqrt{2\pi}}
\left[2\Delta^{*2}-3\left(\frac{1}{2}-\al^2\right)
+\frac{3}{2}\sqrt{2\pi}\al\Delta^*\right]\right\}.
\end{align}
\end{widetext}

\begin{figure}
{\includegraphics[width=0.85\columnwidth]{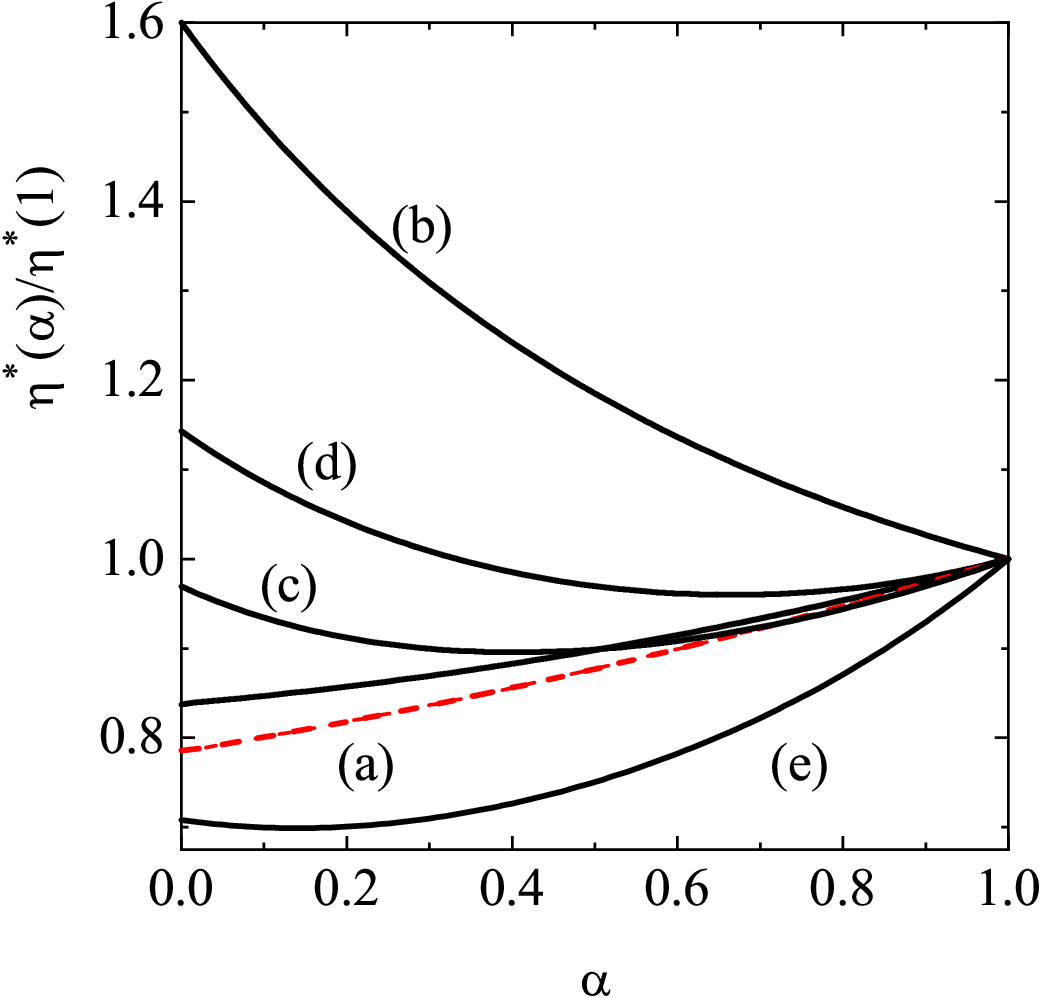}}
\caption{Plot of the (scaled) shear viscosity coefficient $\eta^*(\al)/\eta^*(1)$ as a function of the coefficient of restitution $\al$ for $d=2$ and two values of the solid volume fraction $\phi$: $\phi=0$ (a) and $\phi=0.2$ (dashed line). The lines (b) and (c) correspond to the results obtained in the conventional IHS model for $\phi=0$ (b) and $\phi=0.2$ (c). The lines (d) and (e) correspond to the results obtained in the stochastic heated model for $\phi=0$ (d) and $\phi=0.2$ (e).
\label{fig1}}
\end{figure}

\begin{figure}
{\includegraphics[width=0.85\columnwidth]{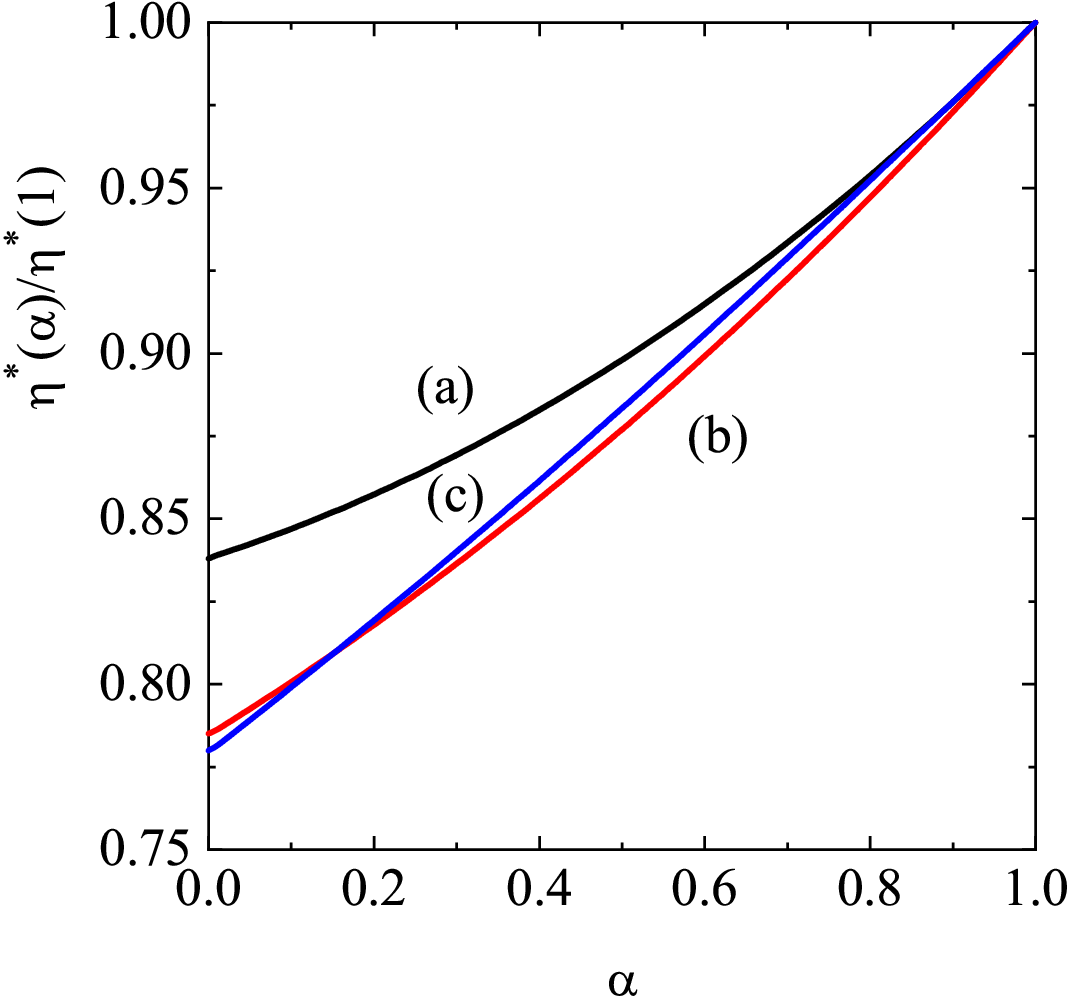}}
\caption{Plot of the (scaled) shear viscosity coefficient $\eta^*(\al)/\eta^*(1)$ as a function of the coefficient of restitution $\al$ for $d=2$ and three values of the solid volume fraction $\phi$: $\phi=0$ (a), $\phi=0.2$ (b), and $\phi=0.4$ (c).
\label{fig2}}
\end{figure}

\begin{figure}
{\includegraphics[width=0.85\columnwidth]{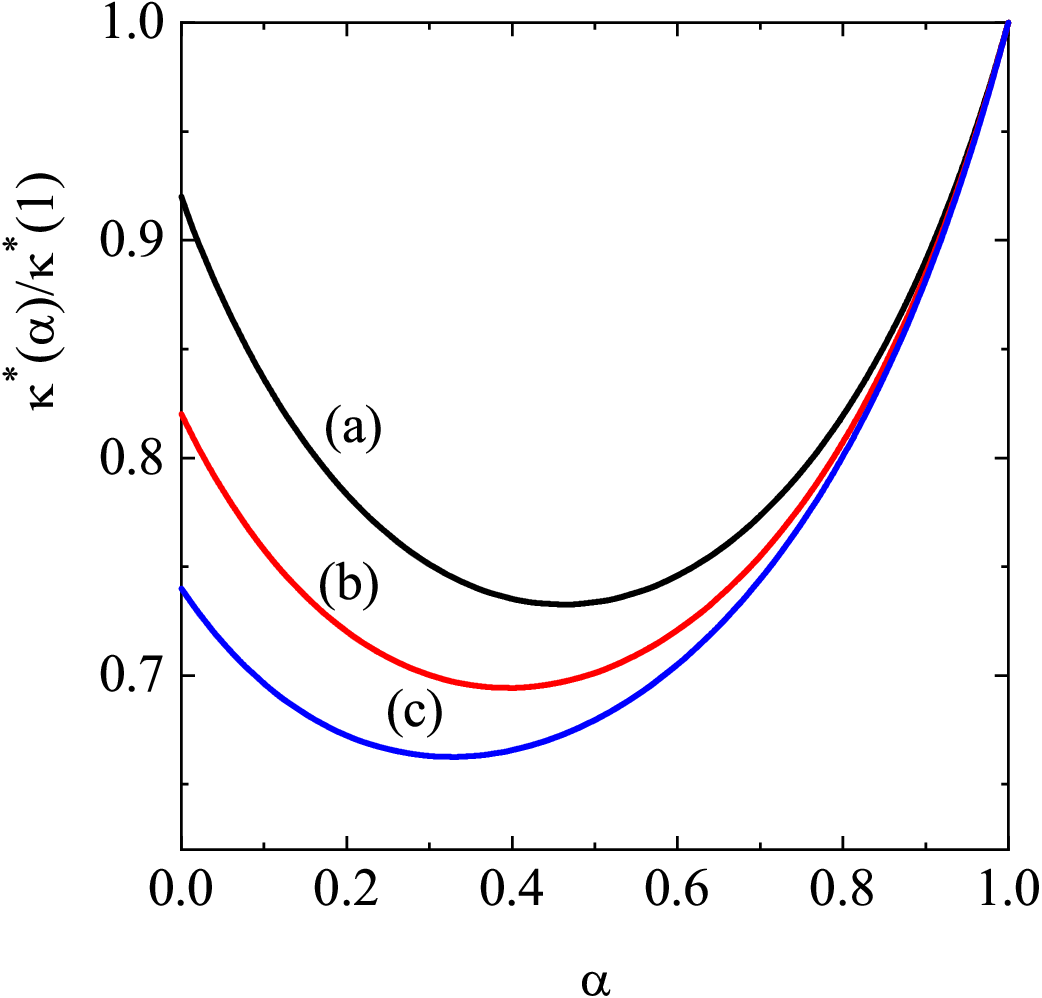}}
\caption{Plot of the (scaled) thermal conductivity coefficient $\kappa^*(\al)/\kappa^*(1)$ as a function of the coefficient of restitution $\al$ for $d=2$ and three values of the solid volume fraction $\phi$: $\phi=0$ (a), $\phi=0.2$ (b), and $\phi=0.4$ (c).
\label{fig3}}
\end{figure}
\begin{figure}
{\includegraphics[width=0.85\columnwidth]{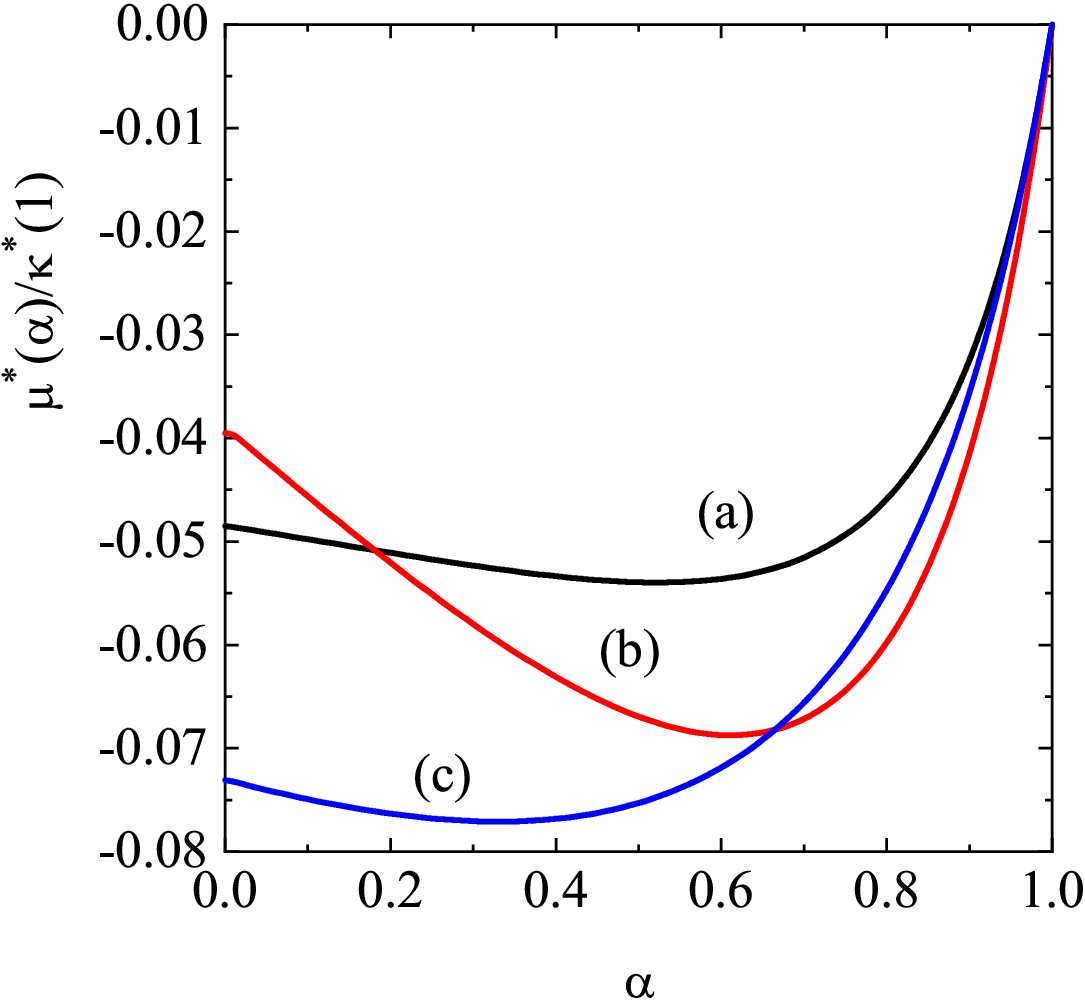}}
\caption{Plot of the (scaled) diffusive heat conductivity coefficient $\mu^*(\al)/\kappa^*(1)$ as a function of the coefficient of restitution $\al$ for $d=2$ and three values of the solid volume fraction $\phi$: $\phi=0$ (a), $\phi=0.2$ (b), and $\phi=0.4$ (c). Here, in contrast to the previous transport coefficients, the curves have been obtained from Table \ref{table1} by adding the pure \emph{dilute} contribution $a_{2,\text{s}}/2\nu_\kappa^*$. The inclusion of this term leads to a nonzero diffusive heat conductivity coefficient when $\phi=0$.
\label{fig4}}
\end{figure}

It is possible now to obtain the transport coefficients for any value of $\phi$, $\alpha$, and $\Delta^*$ as numerical solutions of the  differential equations \eqref{3.9}, \eqref{3.20}, and \eqref{3.24}. However, as the temperature  reaches (on a time scale given by the dissipative collisions) local stationary values that depend on the inelasticity, it is more instructive to evaluate the transport coefficients under this condition. The resulting transport coefficients describe the dynamics close to the stationary state. As said before, in the steady state, $\zeta^{(0)}=0$ and $\Delta^*\to \Delta_\text{s}^*$.\; With this result, all the (scaled) Navier--Stokes transport coefficients can be explicitly written in terms of the volume fraction $\phi$ and the coefficient of restitution $\al$.  The results are summarized in Table~\ref{table1}. The transport coefficients have been reduced by $\eta_0$ and $\kappa_0$, namely,
\begin{align}
\label{4.10}
\eta^*&\equiv \eta/\eta_0, \quad \gamma^*\equiv \gamma/\eta_0,\\
\label{4.11}
\kappa^*&\equiv \kappa/\kappa_0, \quad \mu^*\equiv n \mu/T \kappa_0.
\end{align}
Also shown in this table is the approximation for $\chi$ as a function of $\phi$ for disks \cite{JM87}. It is easy to check that all results presented in Table \ref{table1} reduce to the previous ones recently obtained for the $\Delta$-model in Refs.\ \cite{BBMG15,SRB14} in the dilute regime ($\phi=0$).

Figure~\ref{fig1} shows the dependence of the (scaled) shear viscosity coefficient $\eta^*(\al)/\eta^*(1)$ versus the coefficient of restitution $\al$ for two values of the solid volume fraction $\phi$: $\phi=0$ (dilute gas) and $\phi=0.2$ (moderately dense gas). The corresponding results obtained for this coefficient in the conventional IHS model \cite{GD99,L05} and the stochastic heated model \cite{G13bis} are also plotted for the sake of comparison. It is quite apparent first that the scaled shear viscosity coefficient exhibits a weak dependence on the density in the Delta-model in comparison with the density dependence found in the other models. This is quite an unexpected result since there are new contributions to the shear viscosity coming from density effects not accounted for in the previous results \cite{BBMG15} obtained for a dilute gas ($\phi=0$). To show more clearly the density effects on $\eta^*$, the ratio $\eta^*(\al)/\eta^*(1)$ is plotted in Fig.\ \ref{fig2} as a function of $\al$ for three different values of the solid volume fraction. We observe that,  except from extreme values of dissipation, the shear viscosity (scaled with respect to its elastic value) decreases with density. On the other hand, a different behavior is found for the (scaled) thermal conductivity $\kappa^*(\al)/\kappa^*(1)$. This illustrated in Fig.\ \ref{fig3} where we find that the (scaled) thermal conductivity $\kappa^*(\al)/\kappa^*(1)$ exhibits a non-monotonic dependence on the coefficient of restitution. In addition, it is also quite apparent that the effect of both dissipation and density on the thermal conductivity is more significant that the one observed for the shear viscosity.

Finally, the dependence of the heat diffusive coefficient $\mu^*$ on $\al$ is shown in Fig.\ \ref{fig4} for three different densities. Note that the coefficient $\mu^*$ vanishes in the low-density regime ($\phi=0$) when one neglects the contribution of the kurtosis $a_{2,\text{s}}$ since $\mu^* \propto a_{2,\text{s}}$ in this regime \cite{BBMG15}. Thus, in order to compare the kinetic contribution to $\mu^*$ when $\phi=0$ with the collisional contributions to this coefficient for dense gases ($\phi\neq 0$), the three curves considered in Fig.\ \ref{fig4} have been obtained from the results displayed in Table \ref{table1} \emph{but} including  the pure \emph{dilute} contribution to $\mu^*$ coming from $a_{2,\text{s}}$. This (new) contribution is given by $a_{2,\text{s}}/2\nu_\kappa^*$ for $d=2$ \cite{BBMG15}. Note also that the collision integral \eqref{4.7} appearing in the expression of $\mu^*$ has been still computed by assuming $a_{2,\text{s}}=0$ for the sake of simplicity. It is quite apparent that this coefficient is always negative and the impact of density on $\mu^*$ is in general quite significant. Moreover, as expected, the magnitude of $\mu^*$ is quite small for any density. This means that, for practical purposes, one can neglect the contribution to the heat flux coming from the term proportional to the density gradient, namely, the heat flux obeys Fourier's law $\mathbf{q}^{(1)}=-\kappa \nabla T$. This conclusion contrasts with the results obtained for undriven granular fluids \cite{GD99} where the (reduced) heat diffusive coefficient $\mu^*$ is clearly different from zero for strong collisional dissipation.


\section{Discussion}
\label{sec5}

In this paper we have derived the expressions for transport coefficients for the so-called Delta-model.
This model is an extension of the usual IHS one, where in every particle collision, an amount  $\Delta$ is added to the velocity of the colliding particles in the  normal direction [see Eqs.~(\ref{1.3}) and (\ref{1.4})]. As a consequence, there is an energy input into the system
and a steady state with homogeneous temperature and density is reached, as opposite to the  IHS model.  In this sense, the Delta-model can be seen as a thermostated system by the energy injection via the $\Delta$ parameter.  The stationary temperature depends on $\alpha$ and $\Delta$,
but it is almost independent of the density, due to the fact that both dissipation and energy injection act via interparticle collisions.

We take as starting point the Enskog kinetic equation for the single particle distribution function, that is
solved perturbatively by a Chapman--Enskog expansion in the spatial gradients of the hydrodynamic fields.
With such scheme, it is possible to calculate the Navier--Stokes
transport coefficients: $\eta$ (shear viscosity that appears in the stress tensor) and $\kappa,\,\mu$
[thermal and diffusive heat conductivities, that enter in the heat flux, Eq.~(\ref{3.10b})].

Results for the scaled shear viscosity $\eta^*$, reported in Fig.~\ref{fig2}, show a weak dependence
with the density.
This is a relatively surprising result since, according to the results displayed in Table \ref{table1}, there are new contributions to $\eta^*$ accounting for density corrections to the momentum transport. These corrections were not of course evaluated in the previous work for dilute gases \cite{BBMG15}. On the other hand, given the intricacies associated with the evaluation of the shear viscosity from the Chapman--Enskog solution, it is difficult to provide an intuitive physical explanation for this weak dependence of $\eta^*$ on density in confined granular systems.\; Therefore, according to the previous finding, the main dependence of $\eta^*$ on the density is via the elastic dimensionless viscosity, $\eta^*(\alpha=1)$.  This weak dependence is consistent with the results of numerical simulations, where the scaled shear viscosities are also similar between dilute \cite{SRB14} and dense ($\phi=0.314$) \cite{BRS13} systems.
On the other hand, the influence of density on the (scaled) thermal conductivity $\kappa^*(\alpha)/\kappa^*(\alpha=1)$ is slightly larger than that of the viscosity (see Fig.\ \ref{fig3}).
Finally, in Fig.~\ref{fig4} we represent the diffusive heat conductivity coefficient,
$\mu^*(\alpha)$, divided by the thermal conductivity for an elastic system, $\kappa^*(\alpha=1)$,
as the coefficient $\mu(\alpha)$ vanishes at $\alpha=1$. Here, in contrast to the previous plots and in order to assess the impact of density on $\mu^*$, the pure dilute contribution $(d-1/d)(a_{2,\text{s}}/\nu_\kappa^*)$ has been also considered in the expression of $\mu^*$. This yields $\mu^*\neq 0$ in the low-density limit ($\phi=0$). We observe that $\mu^*$ is the most
sensitive with respect to the density, probably because it vanishes for elastic fluids, where
there is no heat transport associated with density inhomogeneities. In addition, given that the magnitude of the (scaled) diffusive heat conductivity $\mu^*$ is much smaller than that of the (scaled) thermal conductivity $\kappa^*$, one can neglect the term proportional to the density gradient in the heat flux. Therefore, for practical purposes and analogously to ordinary (elastic) gases, one can assume that the heat flux verifies Fourier's law $\mathbf{q}^{(1)}=-\kappa \nabla T$.


As it has been mentioned through the paper, the calculations of the Navier--Stokes transport coefficients have been obtained by considering the leading terms in a Sonine expansion. In addition, non-Gaussian corrections to the zeroth order distribution function $f^{(0)}$ have been neglected for the sake of simplicity. To discuss the validity of the latter approximation, we can use the results of Ref.\ \cite{SRB14} for dilute gases, where the cumulants $a_i$ of the scaled stationary distribution $\varphi$ were measured by computer simulations. For the whole range of inelasticities, $a_2\lesssim 0.10$, $a_3\lesssim 0.07$, and $a_4\lesssim 0.03$, values which are similar or even smaller than other models, in particular than those reported for IHS \cite{SantosMontanero09}.
In the above work \cite{SRB14}, the shear viscosity  was determined by keeping both higher order Sonine polynomials in the stationary distribution and also in the trial function used to evaluate the shear viscosity. Their conclusion is that one has to keep  Sonine corrections to the same order in both places
to obtain a quantitative agreement with simulations. Otherwise, the predictions for the shear viscosity fail in about 15\%.
Therefore the calculations presented here, that neglect both Sonine corrections, are expected to give the qualitative density dependence of transport coefficients,
but may fail when predicting quantitative results by the aforementioned 15\%.
Expressions for the transport coefficients including additional contributions of Sonine polynomials are cumbersome and will be presented elsewhere.
As Chapman--Enskog method gives very
good predictions for moderate densities in other models, like elastic fluids or IHS, we expect that it will be the case fo the Delta-model when proper Sonine corrections are retained in Chapman--Enskog solution.

A possible application of the results derived in this paper might be the study of the stability of the homogeneous steady state. The stability of this state depends on the stationary transport coefficients obtained here and also on the dynamics of these coefficients in the vicinity of the steady state. The linear hydrodynamic stability of the present confined model has been recently analyzed in the dilute regime \cite{BBGM16}. The stability analysis carried out around the time-dependent homogeneous state shows that in some cases the linear analysis is not sufficient to achieve a conclusion on the stability. On the other hand, MD simulations have confirmed the stability of the time-dependent homogeneous state of the Delta-model \cite{BBGM16}. In the dense regime under study here, the steady state is expected to be also stable. First, the compressibility, derived from the expression of the pressure, is always positive \cite{BRS13} and no van der Waals instability takes place. Second, the linear dynamics around the steady state is governed by the hydrodynamic matrix (see Eq.\ (10) in Ref.\ \cite{BRS13}). Using the transport coefficients computed in this article, one obtains that the hydrodynamic matrix remains positive definite for all densities and inelasticities. We plan to perform in the near future a more careful study on this stability.


\acknowledgments

The research of VG has been supported by the Spanish Government through grant No.\ FIS2016-76359-P, partially financed by FEDER funds.  The research of RB and RS has been supported by the Spanish Government, grant No.~FIS2014-52486-R and FIS2017-83709-R. RS has been supported by the Fondecyt Grant No.\ 1140778.

\vspace{0.5cm}

\appendix

\section{Balance equations. Collisional transfer contributions}
\label{appC}

In this Appendix we provide some technical details on the derivation of the expressions of the cooling rate and the collisional transfer contributions to the pressure tensor and the heat flux.
Using Eq.\ \eqref{1.16}, the identity \eqref {1.9} of the Enskog collision integral can be expressed as
\beqa
\label{c1}
I_\psi &\equiv & \int \dd \mathbf{v}_1 \psi(\mathbf{v}_1) J_{\text{E}}[\mathbf{r},\mathbf{v}_1|f,f]
\nonumber\\
&=&\sigma^{d-1}\int \dd{\bf v}_{1}\int \dd{\bf v}_{2}\int \dd\widehat{\boldsymbol{\sigma}}
\Theta (\widehat{{\boldsymbol {\sigma }}}\cdot {\bf g})(\widehat{\boldsymbol {\sigma }}\cdot {\bf g})
 \nonumber\\
& & \times \left[\psi(\mathbf{v}_1')-\psi(\mathbf{v}_1)\right]
f_2(\mathbf{r},\mathbf{v}_1,\mathbf{r}+\boldsymbol{\sigma},\mathbf{v}_2,t),
\eeqa
where $\psi(\mathbf{v})$ is an arbitrary function of velocity.
Equation \eqref{c1} can be written in an equivalent form by interchanging $\mathbf{v}_1$ and $\mathbf{v}_2$ and changing $\widehat{\boldsymbol {\sigma }}\to -\widehat{\boldsymbol {\sigma }}$ to give
\beqa
\label{c3}
I_\psi
&=&\sigma^{d-1}\int \dd{\bf v}_{1}\int \dd{\bf v}_{2}\int \dd\widehat{\boldsymbol{\sigma}}
\Theta (\widehat{{\boldsymbol {\sigma }}}\cdot {\bf g})(\widehat{\boldsymbol {\sigma}}\cdot {\bf g})
\nonumber\\
& & \times \left[\psi(\mathbf{v}_2')-\psi(\mathbf{v}_2)\right]f_2(\mathbf{r},\mathbf{v}_2,\mathbf{r}-\boldsymbol{\sigma},\mathbf{v}_1,t).
\eeqa
\begin{widetext}
Combination of Eqs.\ \eqref{c1} and \eqref{c3} leads to the identity
\beqa
\label{c4}
I_\psi
&=&\frac{1}{2}\sigma^{d-1}\int \dd{\bf v}_{1}\int \dd{\bf v}_{2}\int \dd\widehat{\boldsymbol{\sigma}}
\Theta (\widehat{{\boldsymbol {\sigma}}}\cdot {\bf g})(\widehat{\boldsymbol {\sigma }}\cdot {\bf g})
\Big\{\left[\psi(\mathbf{v}_1')-\psi(\mathbf{v}_1)\right]
f_2(\mathbf{r},\mathbf{v}_1,\mathbf{r}+\boldsymbol{\sigma},\mathbf{v}_2,t)
\nonumber\\
& & +\left[\psi(\mathbf{v}_2')-\psi(\mathbf{v}_2)\right]
f_2(\mathbf{r},\mathbf{v}_2,\mathbf{r}-\boldsymbol{\sigma},\mathbf{v}_1,t)\Big\}.
\eeqa
To simplify Eq.\ \eqref{c4},  note first the relation
\beq
\label{c5}
f_2(\mathbf{r},\mathbf{v}_2,\mathbf{r}-\boldsymbol{\sigma},\mathbf{v}_1,t)=
f_2(\mathbf{r}-\boldsymbol{\sigma},\mathbf{v}_1,\mathbf{r},\mathbf{v}_2,t)
\eeq
and then arrange terms to achieve the result
\beqa
\label{c6}
I_\psi&=& \frac{1}{2}\sigma^{d-1}\int \dd{\bf v}_{1}\int \dd{\bf v}_{2}\int \dd\widehat{\boldsymbol{\sigma}}
\Theta (\widehat{{\boldsymbol {\sigma }}}\cdot {\bf g})(\widehat{\boldsymbol {\sigma }}\cdot {\bf g})
\left\{\left[\psi(\mathbf{v}_1')+\psi(\mathbf{v}_2')-\psi(\mathbf{v}_1)-\psi(\mathbf{v}_2)\right]
f_2(\mathbf{r},\mathbf{v}_1,\mathbf{r}+\boldsymbol{\sigma},\mathbf{v}_2,t)\right.\nonumber\\
& &\left. +
\left[\psi(\mathbf{v}_1')-\psi(\mathbf{v}_1)\right]
\left[f_2(\mathbf{r},\mathbf{v}_1,
\mathbf{r}+\boldsymbol{\sigma},\mathbf{v}_2,t)-
f_2(\mathbf{r}-\boldsymbol{\sigma},\mathbf{v}_1,\mathbf{r},\mathbf{v}_2,t)\right]
\right\}.
\eeqa
The second term, which  vanishes for dilute gases due to the spatial difference of the colliding pair, can be written as a divergence through the identity
\beqa
\label{c7}
F(\mathbf{r},\mathbf{r}+\boldsymbol{\sigma})-
F(\mathbf{r}-\boldsymbol{\sigma},\mathbf{r})&=&-\int_{0}^{1}\; \dd \lambda \;
\frac{\partial}{\partial \lambda}F[\mathbf{r}-\lambda \boldsymbol{\sigma},\mathbf{r}+
(1-\lambda) \boldsymbol{\sigma}] \nonumber\\
&=& \frac{\partial}{\partial \mathbf{r}}\cdot \boldsymbol{\sigma}
\int_{0}^{1}\; \dd \lambda \;F[\mathbf{r}-\lambda \boldsymbol{\sigma},\mathbf{r}+
(1-\lambda) \boldsymbol{\sigma}],
\nonumber\\
\eeqa
for any function $F$. Using this identity, Eq.\ \eqref{c6} can be rewritten as
\beqa
\label{c8}
I_\psi&=& \frac{1}{2}\sigma^{d-1}\int \dd{\bf v}_{1}\int \dd{\bf v}_{2}\int \dd\widehat{\boldsymbol{\sigma}}
\Theta (\widehat{{\boldsymbol {\sigma }}}\cdot {\bf g})(\widehat{\boldsymbol {\sigma }}\cdot {\bf g})
\left\{\left[\psi(\mathbf{v}_1')+\psi(\mathbf{v}_2')-\psi(\mathbf{v}_1)-\psi(\mathbf{v}_2)\right]
f_2(\mathbf{r},\mathbf{v}_1,\mathbf{r}+\boldsymbol{\sigma},\mathbf{v}_2,t)\right.\nonumber\\
& &+ \left. \nabla \cdot \boldsymbol{\sigma}
\left[\psi(\mathbf{v}_1')-\psi(\mathbf{v}_1)\right]
\int_0^{1} \dd \lambda f_2\left[\mathbf{r}-\lambda \boldsymbol{\sigma},\mathbf{v}_1,
\mathbf{r}+(1-\lambda)\boldsymbol{\sigma},\mathbf{v}_2,t)\right]\right\}.
\nonumber\\
\eeqa
The first term in Eq.\ \eqref{c8}, which  also appears in the case of a dilute gas, represents a collisional effect due to  scattering with a change in the velocities.  The second term provides the collisional transfer contributions to the momentum and heat fluxes.

Equation \eqref{c8} is general since it applies to any scattering model. We consider now the collisional model defined by the collision rules \eqref{1.1} and \eqref{1.2}.
For $\psi=m$, $I_m$ vanishes identically. In the case $\psi(\mathbf{v})=m\mathbf{v}$, the first term in the integrand \eqref{c8} vanishes since the momentum is conserved in all pair collisions, i.e., $\mathbf{v}_1'+\mathbf{v}_2'=\mathbf{v}_1+\mathbf{v}_2$. Thus, Eq.\ \eqref{c8} for $\psi(\mathbf{v})=m\mathbf{v}$ reduces to
\beqa
\label{c9}
I_{m\mathbf{v}}&=&
-\nabla\cdot \frac{1+\alpha}{4}m\sigma^{d}
\int \dd \mathbf{v}_1\int \dd \mathbf{v}_2
\int \dd\widehat{\boldsymbol {\sigma }}\,\Theta (\widehat{{\boldsymbol {\sigma}}}
\cdot \mathbf{g})(\widehat{\boldsymbol {\sigma }}\cdot {\bf g})\widehat{\boldsymbol {\sigma }}
\widehat{\boldsymbol {\sigma }}\left[(\widehat{\boldsymbol {\sigma}}\cdot {\bf g})+\frac{2\Delta}{1+\al}\right]
\nonumber \\
& & \times \int_{0}^{1}\; \dd \lambda \; f_2\left[\mathbf{r}-\lambda \boldsymbol{\sigma},\mathbf{v}_1,
\mathbf{r}+(1-\lambda)\boldsymbol{\sigma},\mathbf{v}_2,t)\right].
\eeqa

According to the momentum balance equation (\ref{1.11}), the divergence of the collisional transfer part $\mathsf{P}_{\text{c}}$ is defined by
\begin{equation}
\label{c10}
I_{m\mathbf{v}}=-\nabla\cdot \mathsf{P}_{\text{c}}.
\end{equation}
The explicit form \eqref{1.14} for $\mathsf{P}_{\text{c}}$ may be easily identified after comparing Eqs.\ \eqref{c8} and \eqref{c10}.

The case of kinetic energy $\psi=\frac{1}{2}mv^2$ can be analyzed in a similar way except that energy is not conserved in collisions. This means that the first term on the right side of Eq.\ \eqref{b6} does not vanish. As before, the second term on the right side of Eq.\ \eqref{b7} gives the collisional transfer contribution to the heat flux. After a simple algebra, one obtains
\beqa
\label{c11}
I_{\frac{m}{2}v^2}&=&\frac{m}{2}\sigma^{d-1}\int \dd \mathbf{v}_1\int \dd \mathbf{v}_2
\int d\widehat{\boldsymbol {\sigma }}\,\Theta (\widehat{{\boldsymbol {\sigma}}}
\cdot \mathbf{g})(\widehat{\boldsymbol {\sigma}}\cdot {\bf g})\left[\Delta^2+\alpha \Delta (\widehat{\boldsymbol {\sigma }}
\cdot {\bf g})-\frac{1-\al^2}{4}(\widehat{\boldsymbol {\sigma }}\cdot {\bf g})^2\right]
f_2(\mathbf{r},\mathbf{v}_1,\mathbf{r}+\boldsymbol{\sigma},\mathbf{v}_2,t)
\nonumber\\
& &-\nabla\cdot m\sigma^{d}\frac{1+\al}{4}\int \dd \mathbf{v}_1\int \dd \mathbf{v}_2
\int \dd\widehat{\boldsymbol {\sigma }}\,\Theta (\widehat{{\boldsymbol {\sigma}}}
\cdot \mathbf{g})(\widehat{\boldsymbol {\sigma}}\cdot {\bf g})^2 \widehat{\boldsymbol {\sigma}}
\left[\frac{1-\al}{4}(\widehat{\boldsymbol {\sigma }}\cdot {\bf g})+\widehat{\boldsymbol {\sigma }}\cdot {\bf G}
+\widehat{\boldsymbol {\sigma }}\cdot {\bf U}\right]
\nonumber\\
& &\times
\int_{0}^{1}\; \dd \lambda \; f_2\left[\mathbf{r}-\lambda \boldsymbol{\sigma},\mathbf{v}_1,
\mathbf{r}+(1-\lambda)\boldsymbol{\sigma},\mathbf{v}_2,t)\right],
\nonumber\\
& & +
\nabla\cdot \frac{m}{4}\sigma^{d}\int \dd \mathbf{v}_1\int \dd \mathbf{v}_2
\int \dd\widehat{\boldsymbol {\sigma }}\,\Theta (\widehat{{\boldsymbol {\sigma}}}
\cdot \mathbf{g})(\widehat{\boldsymbol {\sigma}}\cdot {\bf g}) \widehat{\boldsymbol {\sigma}}
\left[\Delta^2+(1+\al)(\widehat{\boldsymbol {\sigma}}\cdot {\bf g})\Delta -2\Delta (\widehat{\boldsymbol {\sigma }}\cdot \mathbf{V}_1)-2\Delta (\widehat{\boldsymbol {\sigma}}\cdot\mathbf{U})\right]\nonumber\\
& &\times
\int_{0}^{1}\; \dd \lambda \; f_2\left[\mathbf{r}-\lambda \boldsymbol{\sigma},\mathbf{v}_1,
\mathbf{r}+(1-\lambda)\boldsymbol{\sigma},\mathbf{v}_2,t)\right],
\eeqa
where we recall that $\mathbf{G}=(\mathbf{V}_1+\mathbf{V}_2)/2$ and $\mathbf{V}=\mathbf{v}-\mathbf{U}$. Upon deriving Eq.\ \eqref{c11}, use has been made of the relation
\beqa
\label{c12}
v_1^2-v_1^{'2}&=&\frac{1-\al^2}{4}(\widehat{\boldsymbol {\sigma }}\cdot {\bf g})^2+(1+\al)
(\widehat{\boldsymbol {\sigma }}\cdot {\bf g})\left[
(\widehat{\boldsymbol {\sigma }}\cdot {\bf G})
+(\widehat{\boldsymbol {\sigma }}\cdot {\bf U})\right]\nonumber\\
& &
-\Delta^2+2\Delta (\widehat{\boldsymbol {\sigma }}\cdot \mathbf{V}_1)+2\Delta (\widehat{\boldsymbol {\sigma }}\cdot
\mathbf{U})-(1+\al)(\widehat{\boldsymbol {\sigma }}\cdot {\bf g})\Delta.
\eeqa
Moreover, notice that the first contribution in the second term on the right-hand side of Eq.\ \eqref{c11} vanishes by symmetry.  The balance energy equation yields
\begin{equation}
\label{c13}
\int \dd {\bf v} \frac{m}{2} ({\bf v}-{\bf U})^2 J_{\text{E}}[\mathbf{r},{\bf v}|f,f]=-\nabla \cdot {\bf q}_{\text{c}}
-\mathsf{P}_{\text{c}}:\nabla {\bf U}-\frac{d}{2}nT \zeta,
\end{equation}
where ${\bf q}_{\text{c}}$ is the collisional contribution to the heat flux and $\zeta$ is the cooling rate. Comparing Eqs.\ (\ref{c11}) and (\ref{c13}) and taking into account Eq.\ (\ref{c10}), one obtains the expressions (\ref{1.15}) and (\ref{1.17}) for ${\bf q}_{\text{c}}$ and $\zeta$, respectively.

\section{Navier--Stokes collisional transfer contributions to the pressure tensor and heat flux}
\label{appA}

The collisional transfer contributions to the pressure tensor and heat flux are determined from Eqs.\ \eqref{1.14} and \eqref{1.15}, respectively. To first order in gradients, the collisional pressure tensor is
\beqa
\label{a1}
P_{ij,\text{c}}&=&\frac{1+\al}{4}m \sigma^d \chi \int \dd\mathbf{v}_{1}\int \dd\mathbf{v}_{2}\int
\dd\widehat{\boldsymbol {\sigma }}\Theta (\widehat{\boldsymbol
{\sigma }}\cdot
\mathbf{g})(\widehat{\boldsymbol {\sigma }}\cdot \mathbf{g}) \widehat{\boldsymbol {\sigma}}\widehat{\boldsymbol {\sigma }}\left[(\widehat{\boldsymbol {\sigma }}\cdot \mathbf{g})+\frac{2\Delta}{1+\al}\right]\nonumber\\
& \times &
\left[f^{(1)}(\mathbf{V}_1)f^{(0)}(\mathbf{V}_2)+f^{(1)}(\mathbf{V}_2)
f^{(0)}(\mathbf{V}_1)-\frac{1}{2}f^{(0)}(\mathbf{V}_2)
\boldsymbol{\sigma}\cdot \nabla f^{(0)}(\mathbf{V}_1)
+\frac{1}{2}f^{(0)}(\mathbf{V}_1)
\boldsymbol{\sigma}\cdot \nabla f^{(0)}(\mathbf{V}_2)\right]
\nonumber\\
&=&\frac{1+\al}{2}m \sigma^d \chi \int \dd\mathbf{v}_{1}\int \dd\mathbf{v}_{2}\int
\dd\widehat{\boldsymbol {\sigma }}\Theta (\widehat{\boldsymbol
{\sigma }}\cdot
\mathbf{g})(\widehat{\boldsymbol {\sigma }}\cdot \mathbf{g}) \widehat{\boldsymbol {\sigma}}\widehat{\boldsymbol {\sigma }}\left[(\widehat{\boldsymbol {\sigma }}\cdot \mathbf{g})+\frac{2\Delta}{1+\al}\right]\nonumber\\
& \times &
\left[f^{(1)}(\mathbf{V}_1)f^{(0)}(\mathbf{V}_2)+\frac{1}{2}f^{(0)}(\mathbf{V}_1)
\boldsymbol{\sigma}\cdot \nabla f^{(0)}(\mathbf{V}_2)\right]
\nonumber\\
&\equiv &P_{ij,\text{c}}^{(\text{I})}+P_{ij,\text{c}}^{(\text{II})},
\eeqa
where $P_{ij,\text{c}}^{(\text{I})}$ denotes the contribution to $P_{ij,\text{c}}$ computed in the conventional IHS model and $P_{ij,\text{c}}^{(\text{II})}$ refers to the part involving terms proportional to the velocity parameter $\Delta$.

To perform the angular integrals, we need the results \cite{NE98}
\begin{gather}
\label{a1.1}
\int\; \dd\widehat{\boldsymbol{\sigma}}\,
\Theta (\widehat{{\boldsymbol {\sigma }}}\cdot \mathbf{g})(\widehat{\boldsymbol {\sigma }}\cdot \mathbf{g})^k \widehat{\sigma}_i=B_{k+1}g^{k-1} g_{i},\\
\label{a1.2}
\int\; \dd\widehat{\boldsymbol{\sigma}}\,
\Theta (\widehat{{\boldsymbol {\sigma }}}\cdot \mathbf{g})(\widehat{\boldsymbol {\sigma }}\cdot \mathbf{g})^k \widehat{\sigma}_i\widehat{\sigma}_j=\frac{B_{k}}{k+d}g^{k-2}\left(kg_{i}g_{j}+ g^2\delta_{ij}\right),\\
\label{a1.3}
\int\; \dd\widehat{\boldsymbol{\sigma}}\,
\Theta (\widehat{{\boldsymbol {\sigma }}}\cdot \mathbf{g})(\widehat{\boldsymbol {\sigma }}\cdot \mathbf{g})^k \widehat{\sigma}_i\widehat{\sigma}_j\widehat{\sigma}_\ell=\frac{B_{k+1}}
{k+1+d}g^{k-3}\left[(k-1)g_{i}g_{j}g_{\ell}+ g^2\left(g_{i}\delta_{j\ell}+g_{j}\delta_{i\ell}+g_{\ell}\delta_{ij}
\right)\right].
\end{gather}

The quantity $P_{ij,\text{c}}^{(\text{I})}$ is \cite{GD99,L05,G13}
\beqa
\label{a2}
P_{ij,\text{c}}^{(\text{I})}&=&-\frac{B_2}{d+2}(1+\alpha) n \sigma^d \chi
\eta_\text{k} \left(\partial_j U_i+\partial_i U_j-\frac{2}{d}\delta_{ij}
\nabla \cdot \mathbf{U}\right)\nonumber\\
& -&
B_3 \frac{d+1}{4d^2}m n^2\sigma^{d+1}(1+\al)\chi v_0\left(\int \dd \mathbf{c}_1
\int \dd\mathbf{c}_2\varphi(\mathbf{c}_1)\varphi(\mathbf{c}_2)g^*\right)
\left[\frac{d}{d+2}\left(\partial_j U_i+\partial_i U_j-\frac{2}{d}\delta_{ij}
\nabla \cdot \mathbf{U}\right)+\delta_{ij}\nabla \cdot \mathbf{U}\right],
\nonumber\\
\eeqa
where $\eta_\text{k}$ is the kinetic shear viscosity and the coefficients $B_k$ are defined in Eq.\ \eqref{2.18}. The term $P_{ij,\text{c}}^{(\text{II})}$ is given by
\beqa
\label{a3}
P_{ij,\text{c}}^{(\text{II})}&=&\frac{B_1}{d+1}m \sigma^d \chi \Delta
\int \dd\mathbf{v}_{1}\int \dd\mathbf{v}_{2}\;g^{-1}\left(g_i g_j+g^2 \delta_{ij}\right)f^{(1)}(\mathbf{V}_1) f^{(0)}(\mathbf{V}_2)
\nonumber\\
&-&\frac{B_2}{2(d+2)}m \sigma^{d+1} \chi \Delta \partial_\ell U_k
\int \dd\mathbf{v}_{1}\int \dd\mathbf{v}_{2}\;\left(g_i \delta_{j\ell}+
g_j \delta_{i\ell}+g_\ell \delta_{ij}\right)f^{(0)}(\mathbf{V}_1)
\frac{\partial f_0(\mathbf{V}_2)}{\partial V_{2k}}.
\eeqa
The first term of the right hand side of Eq.\ \eqref{a3} involves the first-order distribution $f^{(1)}$. This distribution is given by Eq.\ \eqref{2.19}. By symmetry reasons, the contributions to $P_{ij,\text{c}}^{(\text{II})}$ coming from $f^{(1)}$ only involve the terms proportional to the unknowns $\mathcal {C}_{ij}\propto \eta_\text{k}$ and $\mathcal{D}\propto \zeta_U$, where $\eta_\text{k}$ is the kinetic shear viscosity. Since $\zeta_U$ is expected to be very small \cite{BBMG15}, the coupling between $P_{ij}^{(\text{II})}$ and $\mathcal{D}$ will be neglected here. Moreover, in the leading Sonine approximation,
\beq
\label{a4}
\mathcal{C}_{ij}(\mathbf{V})\to -\frac{\eta_\text{k}}{n T^2}D_{ij}(\mathbf{V})f_\text{M}(\mathbf{V}),
\eeq
where $f_\text{M}(\mathbf{V})$ is the Maxwellian distribution. Thus, the quantity $P_{ij,\text{c}}^{(\text{II})}$ can be explicitly evaluated by considering the approximation \eqref{a4} with the result
\beqa
\label{a5}
P_{ij,\text{c}}^{(\text{II})}&=&-\frac{B_1}{d+1} n\sigma^d \chi \Delta^* \left(\int \dd\mathbf{c}_1
\int \dd\mathbf{c}_2 g^{*-1}g_x^{*2}g_y^{*2}\varphi_\text{M}(\mathbf{c}_1)
\varphi_\text{M}(\mathbf{c}_2)\right)\eta_\text{k} \left(\partial_iU_j+\partial_jU_i-\frac{2}{d}\delta_{ij}\nabla \cdot \mathbf{U}\right)\nonumber\\
&-&\frac{B_2}{2(d+2)}n^2 \sigma^{d+1}m \chi v_0 \Delta^* \left[
\left(\partial_iU_j+\partial_jU_i-\frac{2}{d}\delta_{ij}\nabla \cdot \mathbf{U}\right)+\frac{d+2}{d}\delta_{ij} \nabla \cdot \mathbf{U}\right].
\eeqa
The collisional transfer contributions to the shear viscosity $\eta$ and the bulk viscosity $\gamma$ can be easily identified from Eqs.\ \eqref{a2} and \eqref{a5}.

The collisional transfer contribution to the heat flux to first order in the gradients can be obtained in a similar way. It can be written as
\beq
\label{a6}
q_{i,\text{c}}\equiv q_{i,\text{c}}^{(\text{I})}+q_{i,\text{c}}^{(\text{II})},
\eeq
where $q_{i,\text{c}}^{(\text{I})}$ has been determined in previous papers \cite{GD99,L05,G13} while the quantity $q_{i,\text{c}}^{(\text{II})}$ is proportional to $\Delta$. The first contribution $q_{i,\text{c}}^{(\text{I})}$ is
\beqa
\label{a7}
q_{i,\text{c}}^{(\text{I})}&=&-\frac{3}{2}\frac{B_2}{d+2}n\sigma^d (1+\al)\chi \left(\kappa_\text{k} \partial_i T+\mu_\text{k} \partial_i n\right)-\partial_i T\; \frac{B_3}{8d}\frac{m\sigma^{d+1}}{T}\chi (1+\al) n^2 v_0^3 \nonumber\\
& &\times \left(
\int\ \dd\mathbf{c}_1 \int\dd\mathbf{c}_2 \varphi(\mathbf{c}_1)
\varphi(\mathbf{c}_2)\left[g^{*-1}({\bf g}^*\cdot {\bf G}^*)^{2}+g^*G^{*2}+\frac{3}{2}g^*({\bf g}^*\cdot {\bf G}^*)+\frac{1}{4}g^{*3}\right]\right),
\eeqa
where $\kappa_\text{k}$ and $\mu_\text{k}$ are the kinetic contributions to the thermal conductivity $\kappa$ and the $\mu$ coefficient, respectively. The second contribution $q_{i,\text{c}}^{(\text{II})}$ is given by
\beqa
\label{a8}
q_{i,\text{c}}^{(\text{II})}&=&-\Delta \frac{m \sigma^d}{4}
\int \dd\mathbf{v}_{1}\int \dd\mathbf{v}_{2}\int
\dd\widehat{\boldsymbol {\sigma}}\Theta (\widehat{\boldsymbol{\sigma}}\cdot
\mathbf{g})(\widehat{\boldsymbol {\sigma}}\cdot \mathbf{g})\widehat{\boldsymbol {\sigma}}\left[\Delta +\al (\widehat{\boldsymbol {\sigma}}\cdot \mathbf{g})-2 (\widehat{\boldsymbol {\sigma}}\cdot \mathbf{G})\right]
\nonumber\\
&\times &\left[f^{(1)}(\mathbf{V}_1)f^{(0)}(\mathbf{V}_2)+f^{(1)}(\mathbf{V}_2)
f^{(0)}(\mathbf{V}_1)-\frac{1}{2}f^{(0)}(\mathbf{V}_2)
\boldsymbol{\sigma}\cdot \nabla f^{(0)}(\mathbf{V}_1)
+\frac{1}{2}f^{(0)}(\mathbf{V}_1)
\boldsymbol{\sigma}\cdot \nabla f^{(0)}(\mathbf{V}_2)\right].
\eeqa
Note that if one changes $\widehat{\boldsymbol {\sigma}}\to -\widehat{\boldsymbol {\sigma}}$ and interchanges the role of particles 1 and 2, then the term proportional to the combination $\Delta +\al (\widehat{\boldsymbol {\sigma}}\cdot \mathbf{g})$ in Eq.\ \eqref{a8} vanishes by symmetry. Thus, the quantity $q_{i,\text{c}}^{(\text{II})}$ reads
\beqa
\label{a8.1}
q_{i,\text{c}}^{(\text{II})}&=&\Delta\; \frac{B_1}{d+1}m\sigma^d \chi \int \dd \mathbf{v}_1
\int \dd \mathbf{v}_2\; g^{-1}\left[g_i (\mathbf{g}\cdot \mathbf{G})+g^2 G_i\right]
f^{(1)}(\mathbf{V}_1)f^{(0)}(\mathbf{V}_2)\nonumber\\
& &+ \partial_i T\; \Delta\; B_2 \frac{m\sigma^{d+1}}{2d}\chi
\int \dd \mathbf{v}_1
\int \dd \mathbf{v}_2\; (\mathbf{g}\cdot \mathbf{G})
f^{(0)}(\mathbf{V}_1)\frac{\partial f^{(0)}(\mathbf{V}_2)}{\partial T}.
\eeqa
As in the case of the pressure tensor, the evaluation of the first term on the right hand side of Eq.\ \eqref{a8.1} requires the knowledge of the first-order distribution $f^{(1)}$. By symmetry, the terms of $f^{(1)}$ contributing to $q_{i,\text{c}}^{(\text{II})}$ are $\boldsymbol{\mathcal{A}}$ and $\boldsymbol{\mathcal{B}}$. In the leading Sonine approximation, these terms are \cite{GD99,L05}
\beq
\label{a9}
\boldsymbol{\mathcal{A}}(\mathbf{V})\to -\frac{2}{d+2}\frac{m}{n T^2}\kappa_\text{k} f_\text{M}(\mathbf{V}) \mathbf{S}(\mathbf{V}), \quad
\boldsymbol{\mathcal{B}}(\mathbf{V})\to -\frac{2}{d+2}\frac{m}{T^3}\mu_\text{k} f_\text{M}(\mathbf{V}) \mathbf{S}(\mathbf{V}).
\eeq
With this result, the quantity $q_{i,\text{c}}^{(\text{II})}$ can be finally written as
\beqa
\label{a10}
q_{i,\text{c}}^{(\text{II})}&=&-\frac{8B_1}{d(d+1)(d+2)}n\sigma^d \chi \Delta^* \left(\kappa_\text{k} \partial_i T+\mu_\text{k} \partial_i n\right)
\left(\int \dd\mathbf{c}_1 \int\dd\mathbf{c}_2 \varphi_\text{M}(\mathbf{c}_1)\varphi(\mathbf{c}_2)
g^{*-1}\left[(\mathbf{g}^*\cdot \mathbf{S}^*)(\mathbf{g}^*\cdot \mathbf{G}^*)\right.\right.\nonumber\\
& & \left.+g^{*2}(\mathbf{G}^*\cdot \mathbf{S}^*)\right]\bigg)
-\partial_i T\; \Delta^* \frac{m\sigma^{d+1}}{4d T}B_2\chi n^2 v_0^3\left(\int \dd\mathbf{c}_1 \int\dd\mathbf{c}_2 \;
\varphi(\mathbf{c}_1)(\mathbf{g}^*\cdot \mathbf{G}^*)\frac{\partial}{\partial \mathbf{c}_2}\cdot (\mathbf{c}_2\varphi(\mathbf{c}_2))\right.\nonumber\\
& & \left.
+\Delta^* \int \dd\mathbf{c}_1 \int\dd\mathbf{c}_2 \; \varphi(\mathbf{c}_1)\frac{\partial \varphi(\mathbf{c}_2)}{\partial \Delta^*}(\mathbf{g}^*\cdot \mathbf{G}^*)\right)\nonumber\\
&=&-\frac{8B_1}{d(d+1)(d+2)}n\sigma^d \chi \Delta^* \left(\kappa_\text{k} \partial_i T+\mu_\text{k} \partial_i n\right)
\left(\int \dd\mathbf{c}_1 \int\dd\mathbf{c}_2 \varphi_\text{M}(\mathbf{c}_1)\varphi(\mathbf{c}_2)
g^{*-1}\left[(\mathbf{g}^*\cdot \mathbf{S}^*)(\mathbf{g}^*\cdot \mathbf{G}^*)\right.\right.\nonumber\\
& & \left.\left.+g^{*2}(\mathbf{G}^*\cdot \mathbf{S}^*)\right]\right)
-\partial_i T\; \Delta^* \frac{m\sigma^{d+1}}{4d T}B_2\chi n^2 v_0^3\left(\frac{d}{2}
+\Delta^* \int \dd\mathbf{c}_1 \int\dd\mathbf{c}_2 \; \varphi(\mathbf{c}_1)\frac{\partial \varphi(\mathbf{c}_2)}{\partial \Delta^*}(\mathbf{g}^*\cdot \mathbf{G}^*)\right),
\eeqa
\end{widetext}
where $\mathbf{S}^*(\mathbf{c}_1)$ is defined by Eq.\ \eqref{3.16} and use has been made of the result
\beq
\label{a11}
\int \dd\mathbf{c}_1 \int\dd\mathbf{c}_2 \varphi(\mathbf{c}_1)\varphi(\mathbf{c}_2)c_2^{2}=\frac{d}{2}.
\eeq
In addition, upon deriving Eq.\ \eqref{a10}, the relation \eqref{2.12} has been also employed. The collisional contributions to $\kappa$ and $\mu$ can be easily identified from Eqs.\ \eqref{a7} and \eqref{a10}.


\section{Collision integrals}
\label{appB}

In this Appendix, the collision integrals \eqref{4.5}--\eqref{4.7} involving the operator $\boldsymbol{\mathcal {K}}$ defined by Eq.\ \eqref{2.29} are evaluated for an arbitrary $d$-dimensional system. To perform these integrals, it is first convenient to write the first term in Eq.\ \eqref{2.29} in a different form by changing $\widehat{\boldsymbol {\sigma }}\to -\widehat{\boldsymbol {\sigma }}$ and $\Delta \to -\Delta$. In this case, the operator $\boldsymbol{\mathcal {K}}$ becomes
\beqa
\label{b1}
\mathcal{K}_{i}[X] &=&\sigma^{d}\chi\int \dd \mathbf{v}_{2}\int \dd\widehat{\boldsymbol {\sigma
}}\Theta (\widehat{\boldsymbol {\sigma}} \cdot
\mathbf{g}+2\Delta)\nonumber\\
& &\times (\widehat{\boldsymbol {\sigma }}\cdot
\mathbf{g}+2\Delta)
\widehat{\sigma}_i \alpha^{-2}f^{(0)}(\mathbf{V}_{1}'')X(\mathbf{V}_{2}'')\nonumber\\
& &+\sigma^{d}\chi\int \dd \mathbf{v}_{2}\int \dd\widehat{\boldsymbol {\sigma
}}\Theta (\widehat{\boldsymbol {\sigma}} \cdot
\mathbf{g})(\widehat{\boldsymbol {\sigma }}\cdot
\mathbf{g})
\widehat{\sigma}_i \nonumber\\
& & \times f^{(0)}(\mathbf{V}_{1})X(\mathbf{V}_{2}),
\nonumber\\
\eeqa
where the velocities $(\mathbf{V}_1'', \mathbf{V}_2'')$ are given by Eqs.\ \eqref{1.6} and \eqref{1.7}. Note that the above changes do not
alter the relationship between $(\mathbf{V}_1'', \mathbf{V}_2'')$ and $(\mathbf{V}_1, \mathbf{V}_2)$.

Let us consider the integral
\beq
\label{b3}
I_\Psi\equiv \int \dd\mathbf{v}_1 \Psi (\mathbf{V}_1) \mathcal{K}_i[X(\mathbf{V}_2)],
\eeq
where $\Psi (\mathbf{V}_1)$ is an arbitrary function of velocity. According to Eq.\ \eqref{b1}, the integral $I_\Psi$ is given by
\beqa
\label{b4}
I_\Psi &=&\sigma^{d}\chi\int \dd \mathbf{v}_{1}\int \dd \mathbf{v}_{2}\int \dd\widehat{\boldsymbol {\sigma
}}\Theta (\widehat{\boldsymbol {\sigma}} \cdot
\mathbf{g}+2\Delta)\nonumber\\
& &\times (\widehat{\boldsymbol {\sigma }}\cdot
\mathbf{g}+2\Delta)
\widehat{\sigma}_i \Psi (\mathbf{V}_1) \alpha^{-2}f^{(0)}(\mathbf{V}_{1}'')X(\mathbf{V}_{2}'')\nonumber\\
& &+\sigma^{d}\chi\int \dd \mathbf{v}_{1}\int \dd \mathbf{v}_{2}\int \dd\widehat{\boldsymbol {\sigma
}}\Theta (\widehat{\boldsymbol {\sigma}} \cdot
\mathbf{g})(\widehat{\boldsymbol {\sigma }}\cdot
\mathbf{g})
\widehat{\sigma}_i \Psi (\mathbf{V}_1) \nonumber\\
& & \times f^{(0)}(\mathbf{V}_{1})X(\mathbf{V}_{2}).
\nonumber\\
\eeqa
Now we change variables to integrate over $\mathbf{V}_1''$ and $\mathbf{V}_2''$ instead of $\mathbf{V}_1$ and $\mathbf{V}_2$ in the first term of Eq.\ \eqref{b4}. The Jacobian of the transformation is $\al$, $\widehat{{\boldsymbol {\sigma }}}\cdot {\bf g}=-\al \widehat{{\boldsymbol {\sigma }}}\cdot {\bf g}''-2 \Delta$, and hence the first term of \eqref{b4} can be recast into the form
\begin{multline}
\label{b5}
\sigma^{d}\chi\int \dd \mathbf{v}_{1}''\int \dd \mathbf{v}_{2}''\int \dd\widehat{\boldsymbol {\sigma
}}\Theta (-\widehat{\boldsymbol {\sigma}} \cdot
\mathbf{g}'')\\
\times (-\widehat{\boldsymbol {\sigma }}\cdot
\mathbf{g}'')
\widehat{\sigma}_i \Psi (\mathbf{V}_1) f^{(0)}(\mathbf{V}_{1}'')X(\mathbf{V}_{2}'')\\
=-\sigma^{d}\chi\int \dd \mathbf{v}_{1}''\int \dd \mathbf{v}_{2}''\int \dd\widehat{\boldsymbol {\sigma
}}\Theta (\widehat{\boldsymbol {\sigma}} \cdot
\mathbf{g}'')\\
\times (\widehat{\boldsymbol {\sigma }}\cdot
\mathbf{g}'')
\widehat{\sigma}_i \Psi (\mathbf{V}_1) f^{(0)}(\mathbf{V}_{1}'')X(\mathbf{V}_{2}''),
\end{multline}
where we have performed again the change of variables $\widehat{\boldsymbol {\sigma }}\to -\widehat{\boldsymbol {\sigma }}$ and $\Delta \to -\Delta$ in the second line of Eq.\ \eqref{b5}. Equation \eqref{b5} contains the pre-collisional velocities $(\mathbf{V}_1'',\mathbf{V}_2'')$ and the post-collisional velocities $(\mathbf{V}_1,\mathbf{V}_2)$. Since the transformation $(\mathbf{V}_1'',\mathbf{V}_2'')\to (\mathbf{V}_1,\mathbf{V}_2)$ is equivalent to $(\mathbf{V}_1,\mathbf{V}_2)\to (\mathbf{V}_1',\mathbf{V}_2')$, we can use the dummy variables $(\mathbf{V}_1,\mathbf{V}_2)$ in \eqref{b5} and hence, $\mathbf{V}_1(\mathbf{V}_1'',\mathbf{V}_2'')$ must be relabeled to $\mathbf{V}_1'(\mathbf{V}_1,\mathbf{V}_2)$ where $\mathbf{V}_1'$ is defined by Eq.\ \eqref{1.1}, namely,
\beq
\label{b6}
\mathbf{V}_1'=\mathbf{V}_1-\frac{1}{2}\left(1+\alpha\right)(\widehat{{\boldsymbol {\sigma }}}\cdot \mathbf{g})\widehat{{\boldsymbol {\sigma }}}-\Delta \widehat{{\boldsymbol {\sigma }}}.
\eeq
Consequently, the integral $I_\Psi$ can be rewritten as
\beqa
\label{b7}
I_\Psi&=&-\sigma^{d}\chi\int \dd \mathbf{v}_{1}\int \dd \mathbf{v}_{2}\int \dd\widehat{\boldsymbol {\sigma
}}\Theta (\widehat{\boldsymbol {\sigma}} \cdot
\mathbf{g})\nonumber\\
& &\times (\widehat{\boldsymbol {\sigma }}\cdot
\mathbf{g})
\widehat{\sigma}_i f^{(0)}(\mathbf{V}_{1})X(\mathbf{V}_{2})\left[\Psi (\mathbf{V}_1')-\Psi (\mathbf{V}_1) \right].
\nonumber\\
\eeqa


Let us evaluate first the integral
\beq
\label{b8}
\mathcal{K}_\eta\equiv \int\; \dd \mathbf{V}
D_{ij}(\mathbf{V}) {\cal K}_i\left[\frac{\partial f^{(0)}}{\partial V_j}\right].
\eeq
By using the identity \eqref{b7}, the integral $K_\eta$ becomes
\begin{widetext}
\beq
\label{b9}
\mathcal{K}_\eta=-\chi \sigma^{d}\int \dd \mathbf{V}_1 \int\ \dd \mathbf{V}_{2}\int \dd\widehat{\boldsymbol{\sigma}}\,
\Theta (\widehat{\boldsymbol {\sigma }}\cdot {\bf g})(\widehat{\boldsymbol {\sigma}}\cdot {\bf g})
\widehat{\sigma}_i\; f^{(0)}(\mathbf{V}_1)\frac{\partial f^{(0)}(\mathbf{V}_2)}{\partial V_{2j}}\left[D_{ij}(\mathbf{V}_1')-D_{ij}(\mathbf{V}_1)\right].
\eeq
The scattering rule \eqref{b6} gives
\beq
\label{b10}
D_{ij}(\mathbf{V}_1')-D_{ij}(\mathbf{V}_1)=-m \Delta \left\{V_{1i}\widehat{\sigma}_j+V_{1j}\widehat{\sigma}_i
-\frac{2}{d}(\widehat{\boldsymbol {\sigma }}\cdot {\bf V}_{1})\delta_{ij}-\left[(1+\al)(\widehat{\boldsymbol {\sigma }}\cdot {\bf g})+\Delta\right]\left(\widehat{\sigma}_i\widehat{\sigma}_j-\frac{1}{d}
\delta_{ij}\right)\right\}+\cdots,
\eeq
where the terms independent of $\Delta$ have been omitted on the right hand side of Eq.\ \eqref{b10} for the sake of concreteness. Thus, the integral $K_\eta$ can be split in two parts; one of them already computed \cite{GD99,L05,G13} for $\Delta=0$ (conventional IHS model) and the other part involving terms proportional to the parameter $\Delta$. In this case, the integral $K_\eta$ can be written as
\beq
\label{b11}
\mathcal{K}_\eta=\mathcal{K}_\eta^{(0)}+\mathcal{K}_\eta^{(1)},
\eeq
where
\beq
\label{b12}
\mathcal{K}_\eta^{(0)}=2^{d-2}(d-1)\chi \phi n T (1+\al)(1-3\al),
\eeq
\beqa
\label{b13}
\mathcal{K}_\eta^{(1)}&=&m \chi \sigma^{d}\Delta \int \dd \mathbf{V}_1 \int\ \dd{\bf V}_{2}\int \dd\widehat{\boldsymbol{\sigma}}\,
\Theta (\widehat{{\boldsymbol {\sigma }}}\cdot {\bf g})(\widehat{\boldsymbol {\sigma }}\cdot {\bf g})
\; f^{(0)}(\mathbf{V}_1)\frac{\partial f^{(0)}(\mathbf{V}_2)}{\partial V_{2j}}\left\{\frac{d-2}{d}(\widehat{\boldsymbol {\sigma }}\cdot {\bf V}_{1})\widehat{\sigma}_j+V_{1j}\right.\nonumber\\
& & \left.-
\frac{d-1}{d}\left[(1+\al)(\widehat{\boldsymbol {\sigma }}\cdot {\bf g})+\Delta\right]\widehat{\sigma}_j\right\}.
\eeqa
Let us compute now the integral $\mathcal{K}_\eta^{(1)}$. Integrating by parts, one gets
\beqa
\label{b14}
\mathcal{K}_\eta^{(1)}&=&\frac{d-1}{d}m \chi \sigma^{d}\Delta \int \dd \mathbf{V}_1 \int\ \dd{\bf V}_{2}
\; f^{(0)}(\mathbf{V}_1)f^{(0)}(\mathbf{V}_2)\int \dd\widehat{\boldsymbol{\sigma}}\,
\Theta (\widehat{{\boldsymbol {\sigma }}}\cdot {\bf g})\left[2(\widehat{{\boldsymbol {\sigma }}}\cdot {\bf V}_{1})-2(1+\al)(\widehat{\boldsymbol {\sigma }}\cdot {\bf g})
-\Delta\right]\nonumber\\
&=&\frac{d-1}{d}m \chi \sigma^{d}\Delta \int \dd \mathbf{V}_1 \int\ \dd{\bf V}_{2}
\; f^{(0)}(\mathbf{V}_1)f^{(0)}(\mathbf{V}_2)\left[2B_1 g^{-1} (\mathbf{g}\cdot \mathbf{V}_1)-2B_1(1+\al)g
-B_0 \Delta\right],
\eeqa
where use has been made of the relations \eqref{a1.1}--\eqref{a1.3} in the last result. The integral $K_\eta^{(1)}$ can be rewritten as
\beq
\label{b15}
\mathcal{K}_\eta^{(1)}=2\frac{d-1}{d}B_1\chi n\sigma^d \Delta^* n T I_\eta'-2^d (d-1)\chi \phi \Delta^{*2}n T,
\eeq
where the dimensionless integral $I_\eta'$ is defined by Eq.\ \eqref{3.9}. This integral can be explicitly estimated by replacing $\varphi$ by its Gaussian form $\varphi_\text{M}=\pi^{-d/2}e^{-c^2}$ with the result
\beq
\label{b16}
I_\eta'=-\sqrt{2}\frac{\Gamma\left(\frac{d+1}{2}\right)}
{\Gamma\left(\frac{d}{2}\right)}(1+2\al).
\eeq
Using of this in Eq.\ \eqref{b15} gives the result
\beq
\label{b17}
\mathcal{K}_\eta^{(1)}=-2^{d}(d-1)\phi \chi \Delta^* n T \left(\sqrt{\frac{2}{\pi}}(1+2\al)
+\Delta^*\right).
\eeq
The expression of $\mathcal{K}_\eta$ can be finally written as
\beq
\label{b18}
\mathcal{K}_\eta=2^{d}(d-1)\phi \chi n T \left[\frac{1}{4}(1+\al)(1-3\al)-\Delta^*
\left(\sqrt{\frac{2}{\pi}}(1+2\al)+\Delta^*\right)\right].
\eeq

We consider now the collision integral
\beqa
\label{b19}
\mathcal{K}_\kappa &\equiv & \int\; \dd \mathbf{V}
\mathbf{S}(\mathbf{V}) \cdot \boldsymbol{{\cal K}}\left[\frac{\partial}{\partial V_i}\left(V_i f^{(0)}\right)
\right]=
-\chi \sigma^{d}\int \dd \mathbf{V}_1 \int\ \dd{\bf V}_{2}
f^{(0)}(\mathbf{V}_1)\frac{\partial}{\partial V_{2i}}\left(V_{2i} f^{(0)}(\mathbf{V}_2)\right)\nonumber\\
& &\times
\int \dd\widehat{\boldsymbol{\sigma}}\,
\Theta (\widehat{{\boldsymbol {\sigma }}}\cdot {\bf g})(\widehat{\boldsymbol {\sigma }}\cdot {\bf g})
\widehat{\sigma}_j\; \left[S_{j}(\mathbf{V}_1')-S_{j}(\mathbf{V}_1)\right].
\eeqa
As in the previous calculation for $K_\eta$, $K_\kappa$ can be divided in two parts, i.e., $\mathcal{K}_\kappa=\mathcal{K}_\kappa^{(0)}+\mathcal{K}_\kappa^{(1)}$. The integral $\mathcal{K}_\kappa^{(0)}$ was evaluated in the case $\Delta=0$ with the result \cite{GD99,L05,G13}
\beq
\label{b20}
\mathcal{K}_\kappa^{(0)}=-\frac{3}{8}2^d d \frac{nT^2}{m}\phi  \chi(1+\alpha)^2\left(2\alpha-1\right),
\eeq
where we have neglected the contribution coming from the kurtosis $a_2$. The other contribution $\mathcal{K}_\kappa^{(1)}$ comes from the terms proportional to $\Delta$. In particular, the contributions to the combination $\widehat{\sigma}_j\; \left[S_{j}(\mathbf{V}_1')-S_{j}(\mathbf{V}_1)\right]$ proportional to $\Delta$ in Eq.\ \eqref{b19} are given by
\beqa
\label{b21}
\widehat{\sigma}_j\; \left[S_{j}(\mathbf{V}_1')-S_{j}(\mathbf{V}_1)\right]&\to & -\frac{m}{2}\Delta \left[ V_1^2
+2 (\widehat{{\boldsymbol {\sigma }}}\cdot {\bf V}_{1})^2-3(1+\al)(\widehat{{\boldsymbol {\sigma }}}\cdot {\bf g})(\widehat{{\boldsymbol {\sigma }}}\cdot {\bf V}_{1})+\frac{3}{4}(1+\al)^2(\widehat{{\boldsymbol {\sigma }}}\cdot {\bf g})^2 \right.\nonumber\\
& & \left. - 3 \Delta (\widehat{{\boldsymbol {\sigma }}}\cdot {\bf V}_{1})+\frac{3}{2}(1+\al)\Delta (\widehat{{\boldsymbol {\sigma }}}\cdot {\bf g})+\Delta^2-\frac{d+2}{m}T\right].
\eeqa
The integral $\mathcal{K}_\kappa^{(1)}$ can be written more explicitly when one takes into account the relation \eqref{b21}. The result is
\beqa
\label{b22}
\mathcal{K}_\kappa^{(1)}&=&\frac{m}{2}\chi \sigma^{d}\Delta \int \dd \mathbf{V}_1 \int\ \dd{\bf V}_{2}
f^{(0)}(\mathbf{V}_1)f^{(0)}(\mathbf{V}_2)\int \dd\widehat{\boldsymbol{\sigma}}\,
\Theta (\widehat{{\boldsymbol {\sigma}}}\cdot {\bf g})(\widehat{\boldsymbol {\sigma }}\cdot {\bf V}_{2})
\nonumber\\
& \times & \left[V_1^2+2(\widehat{{\boldsymbol {\sigma }}}\cdot {\bf V}_{1})^2-6(1+\al)(\widehat{{\boldsymbol {\sigma }}}\cdot {\bf g})(\widehat{{\boldsymbol {\sigma }}}\cdot {\bf V}_{1})+\frac{9}{4}(1+\al)^2(\widehat{{\boldsymbol {\sigma }}}\cdot {\bf g})^2-3\Delta (\widehat{{\boldsymbol {\sigma }}}\cdot {\bf V}_{1})\right.\nonumber\\
& & \left.+3(1+\al)\Delta (\widehat{{\boldsymbol {\sigma }}}\cdot {\bf g})+\Delta^2-\frac{d+2}{m}T\right]
\nonumber\\
&=&\frac{m}{2}\chi \sigma^{d}\Delta \int \dd \mathbf{V}_1 \int\ \dd{\bf V}_{2}
f^{(0)}(\mathbf{V}_1)f^{(0)}(\mathbf{V}_2)\left\{B_1 g^{-1}(\mathbf{g}\cdot \mathbf{V}_2)V_1^2
\right.\nonumber\\
& &
+\frac{2B_1}{d+1}g^{-3}\left[-(\mathbf{g}\cdot \mathbf{V}_1)^2(\mathbf{g}\cdot \mathbf{V}_2)
+g^2\left(2(\mathbf{V}_1\cdot \mathbf{V}_2)(\mathbf{g}\cdot \mathbf{V}_1)+V_1^2(\mathbf{g}\cdot \mathbf{V}_2)\right)\right]\nonumber\\
& &-6(1+\al)\frac{B_1}{d+1}g^{-1}\left[(\mathbf{g}\cdot \mathbf{V}_1)(\mathbf{g}\cdot \mathbf{V}_2)
+g^2(\mathbf{V}_{1}\cdot \mathbf{V}_2)\right]+\frac{9}{4}(1+\al)^2B_3 g(\mathbf{g}\cdot \mathbf{V}_2)
\nonumber\\
& & \left. +3(1+\al)\Delta B_2 (\mathbf{g}\cdot \mathbf{V}_2)+\Delta^2 B_1 g^{-1}(\mathbf{g}\cdot \mathbf{V}_2)
-\frac{d+2}{m}T B_1 g^{-1}(\mathbf{g}\cdot \mathbf{V}_2)\right\}.
\eeqa
As before, this integral is now estimated by replacing $f^{(0)}(\mathbf{V})$ by its Gaussian form $f_\text{M}(\mathbf{V})=n v_0^{-d}\pi^{-d/2} e^{-c^2}$. The result is
\beq
\label{b23}
\mathcal{K}_\kappa^{(1)}=\frac{2^{d-1/2}d}{\sqrt{\pi}}\frac{nT^2}{m}\phi \chi \Delta^*\left[\frac{3}{4}+3(1+\al)\left(1-\frac{1}{2}\sqrt{2\pi}\Delta^*\right)
-\frac{9}{2}(1+\al)^2-\Delta^{*2}\right].
\eeq
With this result $\mathcal{K}_\kappa$ can be written as
\beq
\label{b24}
\mathcal{K}_\kappa=2^d d \frac{n T^2}{m}\phi \chi \left\{\frac{3}{8}(1+\al)^2(1-2\al)+
\frac{1}{\sqrt{2\pi}}\Delta^*\left[\frac{3}{4}+3(1+\al)\left(1-\frac{1}{2}\sqrt{2\pi}\Delta^*\right)
-\frac{9}{2}(1+\al)^2-\Delta^{*2}\right]\right\}.
\eeq

The last integral needed to evaluate the transport coefficient $\mu$ is
\beq
\label{b25}
\mathcal{K}_\mu \equiv  \int\; \dd \mathbf{V}
\mathbf{S}(\mathbf{V}) \cdot \boldsymbol{{\cal K}}\left[f^{(0)}\right]= K_\mu^{(0)}+K_\mu^{(1)},
\eeq
where $\mathcal{K}_\mu^{(0)}$ is \cite{GD99,L05,G13}
\beq
\label{b26}
\mathcal{K}_\mu^{(0)}=-\frac{3}{8}2^d d \frac{nT^2}{m}\phi  \chi \alpha(1-\al^2),
\eeq
and
\beqa
\label{b27}
\mathcal{K}_\mu^{(1)}&=&\frac{m}{2}\chi \sigma^{d}\Delta \int \dd \mathbf{V}_1 \int\ \dd{\bf V}_{2}
f^{(0)}(\mathbf{V}_1)f^{(0)}(\mathbf{V}_2)\int \dd\widehat{\boldsymbol{\sigma}}\,
\Theta (\widehat{{\boldsymbol {\sigma }}}\cdot {\bf g})(\widehat{\boldsymbol {\sigma }}\cdot \mathbf{g})
\nonumber\\
& \times & \left[V_1^2+2(\widehat{{\boldsymbol {\sigma }}}\cdot {\bf V}_{1})^2-3(1+\al)(\widehat{{\boldsymbol {\sigma }}}\cdot {\bf g})(\widehat{{\boldsymbol {\sigma }}}\cdot {\bf V}_{1})+\frac{3}{4}(1+\al)^2(\widehat{{\boldsymbol {\sigma }}}\cdot {\bf g})^2-3\Delta (\widehat{{\boldsymbol {\sigma }}}\cdot {\bf V}_{1})\right.\nonumber\\
& & \left.+\frac{3}{2}(1+\al)\Delta (\widehat{{\boldsymbol {\sigma }}}\cdot {\bf g})+\Delta^2-\frac{d+2}{m}T\right].
\eeqa
The evaluation of $\mathcal{K}_\mu^{(1)}$ follows similar steps as those made before for $K_\kappa^{(1)}$. After a tedious algebra, the expression of $\mathcal{K}_\mu^{(1)}$ can be written as
\beq
\label{b28}
\mathcal{K}_\mu^{(1)}=\frac{2^{d-1/2}d}{\sqrt{\pi}}\frac{nT^2}{m}\phi \chi \Delta^*\left[2\Delta^{*2}-3\left(\frac{1}{2}-\al^2\right)
+\frac{3}{2}\sqrt{2\pi}\al\Delta^*\right].
\eeq
With this expression, the integral $\mathcal{K}_\mu$ is
\beq
\label{b29}
\mathcal{K}_\mu=-2^d d \frac{n T^2}{m}\phi \chi \left\{\frac{3}{8}\al (1-\al^2)-
\frac{1}{\sqrt{2\pi}}\Delta^*\left[2\Delta^{*2}-3\left(\frac{1}{2}-\al^2\right)
+\frac{3}{2}\sqrt{2\pi}\al\Delta^*\right]\right\}.
\eeq
\end{widetext}


\end{document}